\documentclass[preprintnumbers,floatfix,superscriptaddress,nofootinbib,prx,twocolumn,showpacs]{revtex4-2}
\usepackage{amsmath,amsthm,amsfonts,stmaryrd,amssymb}
\usepackage[english]{babel} 
\usepackage[utf8]{inputenc} 
\usepackage[T1]{fontenc} 

\usepackage{fancyhdr}
\usepackage{xcolor}
\usepackage{braket}
\usepackage{enumitem}
\usepackage{multirow}
\usepackage{lipsum}
\usepackage{changepage}
\usepackage{physics}
\usepackage{wasysym}
\usepackage{dsfont}
\usepackage{pifont}
\usepackage{microtype} 

\newcommand{\ssec}[1]{\textls{\textsc{\textbf{#1}}}}

\usepackage{array}
\newcolumntype{L}[1]{>{\raggedright\arraybackslash}m{#1}}
\newcolumntype{C}[1]{>{\centering\arraybackslash}m{#1}}
\newcolumntype{R}[1]{>{\raggedleft\arraybackslash}m{#1}}
\usepackage{floatrow}
\newfloatcommand{capbtabbox}{table}[][\FBwidth]

\usepackage{xspace}

\usepackage{empheq}
\usepackage{siunitx}
\usepackage{mathtools}

\newcommand{\defeq}{\vcentcolon=}
\theoremstyle{definition}

\newtheorem{definition}{Definition}

\theoremstyle{plain}
\newtheorem{lemma}{Lemma}

\usepackage{bm} 

\usepackage{booktabs}
\usepackage{graphicx}
\usepackage{subcaption}
\usepackage{caption}
\usepackage{ragged2e}
\DeclareCaptionJustification{justified}{\justifying}
\usepackage{hyperref}
\hypersetup{
    colorlinks=true,
    linkcolor=blue,
    filecolor=magenta,      
    urlcolor=cyan,
    citecolor=magenta
}

\usepackage{newfloat}
\usepackage{tikz}
\usepackage{framed}

\DeclareFloatingEnvironment[listname=Box,name=Box]{story}
\newlength{\leftybarwidth}
\newlength{\leftybarsep}
\newlength{\leftybarsepo}
\colorlet{leftybarcolor}{black}

\usepackage{blindtext}
\usepackage{nameref}
\newcounter{mylabelcounter}
\makeatletter
\newcommand{\labelText}[2]{%
#1\refstepcounter{mylabelcounter}%
\immediate\write\@auxout{%
  \string\newlabel{#2}{{1}{\thepage}{{\unexpanded{#1}}}{mylabelcounter.\number\value{mylabelcounter}}{}}%
}%
}
\makeatother

\usepackage{algorithm}
\usepackage{algpseudocode}
\makeatletter
\def\BState{\State\hskip-\ALG@thistlm}
\makeatother

\usepackage[utf8]{inputenc}
\usepackage{tabularx}
\usepackage{xspace}
\usepackage{graphicx}
\usepackage{mathtools,amsmath,amsfonts}
\usepackage{dsfont}
\usepackage{tikz}
\usepackage[hmargin=1.5cm, vmargin=2cm]{geometry} 
\graphicspath{{./figs/}}

\newcommand\tc{t_{\mathrm{cut}}}

\allowdisplaybreaks

\begin{document}
\title{Optimal entanglement distribution policies in homogeneous repeater chains with cutoffs}
\date{\today}

\author{Álvaro G. Iñesta}\email{a.gomezinesta@tudelft.nl}
\affiliation{QuTech, Delft University of Technology, Lorentzweg 1, 2628 CJ Delft, The Netherlands}
\affiliation{EEMCS, Delft University of Technology, Mekelweg 4, 2628 CD Delft, The Netherlands}
\affiliation{Kavli Institute of Nanoscience, Delft University of Technology, Lorentzweg 1, 2628 CJ Delft, The Netherlands}
\author{Gayane Vardoyan}\email{g.s.vardoyan@tudelft.nl}
\affiliation{QuTech, Delft University of Technology, Lorentzweg 1, 2628 CJ Delft, The Netherlands}
\affiliation{EEMCS, Delft University of Technology, Mekelweg 4, 2628 CD Delft, The Netherlands}
\author{Lara Scavuzzo}
\affiliation{EEMCS, Delft University of Technology, Mekelweg 4, 2628 CD Delft, The Netherlands}
\author{Stephanie Wehner}
\affiliation{QuTech, Delft University of Technology, Lorentzweg 1, 2628 CJ Delft, The Netherlands}
\affiliation{EEMCS, Delft University of Technology, Mekelweg 4, 2628 CD Delft, The Netherlands}
\affiliation{Kavli Institute of Nanoscience, Delft University of Technology, Lorentzweg 1, 2628 CJ Delft, The Netherlands}

\begin{abstract}
We study the limits of bipartite entanglement distribution using a chain of quantum repeaters that have quantum memories. To generate end-to-end entanglement, each node can attempt the generation of an entangled link with a neighbor, or perform an entanglement swapping measurement. A maximum storage time, known as cutoff, is enforced on the memories to ensure high-quality entanglement. Nodes follow a policy that determines when to perform each operation. Global-knowledge policies take into account all the information about the entanglement already produced. Here, we find global-knowledge policies that minimize the expected time to produce end-to-end entanglement. Our methods are based on Markov decision processes and value and policy iteration. We compare optimal policies to a policy in which nodes only use local information. We find that the advantage in expected delivery time provided by an optimal global-knowledge policy increases with increasing number of nodes and decreasing probability of successful swapping.
Our work sheds light on how to distribute entangled pairs in large quantum networks using a chain of intermediate repeaters with cutoffs.
\end{abstract}

\maketitle

\section{Introduction}\label{sec.intro}
Bipartite entangled states shared between two parties are often required as a basic resource in quantum network applications.
As an example, in cryptography, bipartite entanglement can be directly used for quantum key distribution between two parties \cite{Ekert1991,Bennett1992}, but also in multi-party applications such as quantum secret sharing \cite{BenOr2006}.
Bipartite entanglement can also be used to generate multipartite entangled states that are necessary for other applications \cite{Pirker2018,Kruszynska2006,Bugalho2021}.
As a consequence, a reliable method to distribute entanglement in a quantum network is crucial for the implementation of quantum cryptography applications.

\begin{figure}[t]
	\centering
	\includegraphics[width=0.9\textwidth]{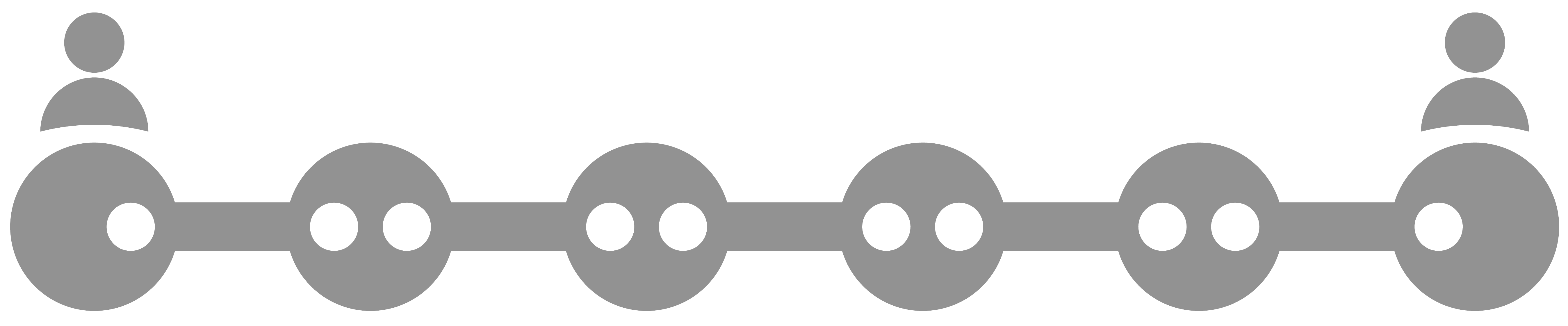}
	\caption{\textbf{A quantum repeater chain that can store two qubits per intermediate node and one qubit per end node.}
	White circles represent qubits. All nodes are equidistant and identical.}
	\label{fig.repeaterChain}
\end{figure}

Two neighboring nodes in a quantum network can generate a shared bipartite entangled state, which we call an entangled link.
This can be done, e.g., by generating an entangled pair at one node and sending half of the pair to the neighbor via an optical fiber \cite{Yoshino2013,Stephenson2020} or free space \cite{Ursin2007,Sidhu2021}. 
Two distant nodes can generate an entangled link by generating entanglement between each pair of adjacent nodes along a path that connects them, and then combining these entangled links into longer-distance bipartite entanglement via entanglement swap operations \cite{Duan2001,Sangouard2011}.
This path constitutes a quantum repeater chain (see Figure~\ref{fig.repeaterChain}).
We consider repeater chains in which nodes can store quantum states in the form of qubits and perform operations and measurements on them.
Experimentally, qubits can be realized with different technologies, such as NV centers \cite{Bernien2013, Hensen2015, Humphreys2018, Rozpedek2019, Pompili2021} and trapped ions \cite{Moehring2007, Slodicka2013}.

We focus on a single repeater chain of $n$ equidistant and identical nodes, which could be part of a larger quantum network.
To generate an entangled link between the two end nodes, also called end-to-end entanglement, we assume the nodes can perform the following operations:
($i$) heralded generation of entanglement between neighbors \cite{Barrett2005,Bernien2013}, which succeeds with probability $p$ and otherwise raises a failure flag;
($ii$) entanglement swaps \cite{Zukowski1993, Duan2001, Sangouard2011},
which consume two adjacent entangled links to generate a longer-distance link
with probability $p_\text{s}$;
and ($iii$) removal of any entangled link that existed for longer than some cutoff time $\tc$, to prevent generation of low-quality end-to-end entanglement due to decoherence \cite{Collins2007, Rozpedek2018, Khatri2019, Rozpedek2019, Li2020}.
Note that cutoff times are a key ingredient, since many applications require quantum states with a high enough quality.

We assume that nodes always attempt entanglement generation if there are qubits available. Cutoffs are always applied whenever an entangled link becomes too old.
However, nodes are free to attempt swaps as soon as entangled links are available or some time later, so they must agree on an entanglement distribution policy: a set of rules that indicate when to perform a swap.
We define an optimal policy as a policy that minimizes the expected entanglement delivery time, which is the average time required to generate end-to-end entanglement.
Here, we consider optimal global-knowledge policies, in which nodes have information about all the entangled links in the chain.
A policy is local when the nodes only need to know the state of the qubits they hold.
An example of local policy is the swap-asap policy, in which each node performs a swap as soon as both entangled links are available.

Previous work on quantum repeater chains has mostly focused on the analysis of specific policies rather than on the search for optimal policies.
For example, \cite{Coopmans2021} provides analytical bounds on the delivery time of a ``nested'' policy \cite{Briegel1998}, and \cite{Jiang2007} optimizes the parameters of such a policy with a dynamic programming approach.
Delivery times can be studied using Markov models.
In \cite{Shchukin2019}, the authors introduce a methodology based on Markov chains to calculate the expected delivery time in repeater chains that follow a particular policy.
Similar techniques have also been applied to other quantum network topologies, such as the quantum switch \cite{Vardoyan2021,Vardoyan2021b}.
Here, we focus on Markov decision processes (MDPs), which have already been applied to related problems, e.g., in \cite{Khatri2022}, the authors use an MDP formulation to maximize the quality of the entanglement generated between two neighboring nodes and between the end nodes in a three-node repeater chain.
Our work builds on \cite{Shchukin2021}, wherein the authors find optimal policies for quantum repeater chains with perfect memories. Since quantum memories are expected to be noisy, particularly in the near future, quantum network protocols must be suitable for imperfect memories. Here, we take a crucial step towards the design of high-quality entanglement distribution policies for noisy hardware. By formulating a generalized MDP to include finite storage times, we are able to find optimal policies in quantum repeater chains with imperfect memories. Our optimal policies provide insights for the design of entanglement distribution protocols.

Our main contributions are as follows:
\begin{itemize}
	\item We introduce a general MDP model for homogeneous repeater chains with memory cutoffs.
	The latter constraint poses a previously unaddressed challenge: MDP states must incorporate not only entangled link absence/presence, but also link age;
	\item We find optimal policies for minimizing the expected end-to-end entanglement delivery time, by solving the MDP via value and policy iteration;
	\item Our optimal policies take into account global knowledge of the state of the chain and therefore constitute a lower bound to the expected delivery time of policies that use only local information.
\end{itemize}
Our main findings are as follows:
\begin{itemize}
	\item The optimal expected delivery time in a repeater chain with deterministic swaps ($p_\mathrm{s}=1$) can be orders of magnitude smaller than with probabilistic swaps;
	\item When swaps are deterministic, the advantage in expected delivery time offered by an optimal policy as compared to the swap-asap policy increases for lower probability of entanglement generation, $p$, and lower cutoff time, $\tc$, in the parameter region explored. However, when swaps are probabilistic, we find the opposite behavior: the advantage increases for higher $p$ and $\tc$;
	\item The advantage provided by optimal policies increases with higher number of nodes, both when swaps are deterministic and probabilistic, albeit the advantage is larger in case of the latter.
\end{itemize}

This paper is structured as follows.
In Section \ref{sec.results}, we explain in detail our repeater chain model and then present our main results.
In Section \ref{sec.discussion}, we discuss the implications and limitations of our work.
In Section \ref{sec.methods}, we provide more details on how to formulate the MDP and how to solve it.

\section{Results}\label{sec.results}
\subsection{Network model}\label{subsec.model}
We analyze quantum repeater chains wherein nodes can store quantum states in the form of qubits and can perform three basic operations with them: entanglement generation, entanglement swaps, and cutoffs.
\\

\ssec{Entanglement generation.} Two adjacent nodes can attempt the heralded generation of an entangled link (i.e., a shared bipartite entangled state), succeeding with probability $p$. Generation of entanglement is heralded, meaning that the nodes receive a message stating whether they successfully generated an entangled link or not \cite{Barrett2005,Bernien2013}.
We assume that entanglement generation is noisy. Hence, the newly generated entangled links are not maximally entangled states but Werner states \cite{Werner1989}. Werner states are maximally entangled states that have been subjected to a depolarizing process, which is a worst-case noise model \cite{Dur2005}, and they can be written as follows:
\begin{equation}
    \rho = \frac{4F-1}{3}\ketbra{\phi^+} + \frac{1-F}{3} \mathbb{I}_4,
\end{equation}
where $\ket{\phi^+}=\frac{\ket{00}+\ket{11}}{\sqrt{2}}$ is a maximally entangled state, $F$ is the fidelity of the Werner state to the state $\ket{\phi^+}$, and $\mathbb{I}_d$ is the $d$-dimensional identity.
In our notation, the fidelity of a mixed state $\rho$ to a pure state $\ket{\phi}$ is defined as
\begin{equation}
    F(\rho,\ket{\phi}) \defeq \bra{\phi}\rho\ket{\phi}.
\end{equation}
We assume that the fidelity of newly generated entangled links is $F_\text{new}\leq1$.
\\

\begin{figure}[t]
	\includegraphics[width=0.65\textwidth]{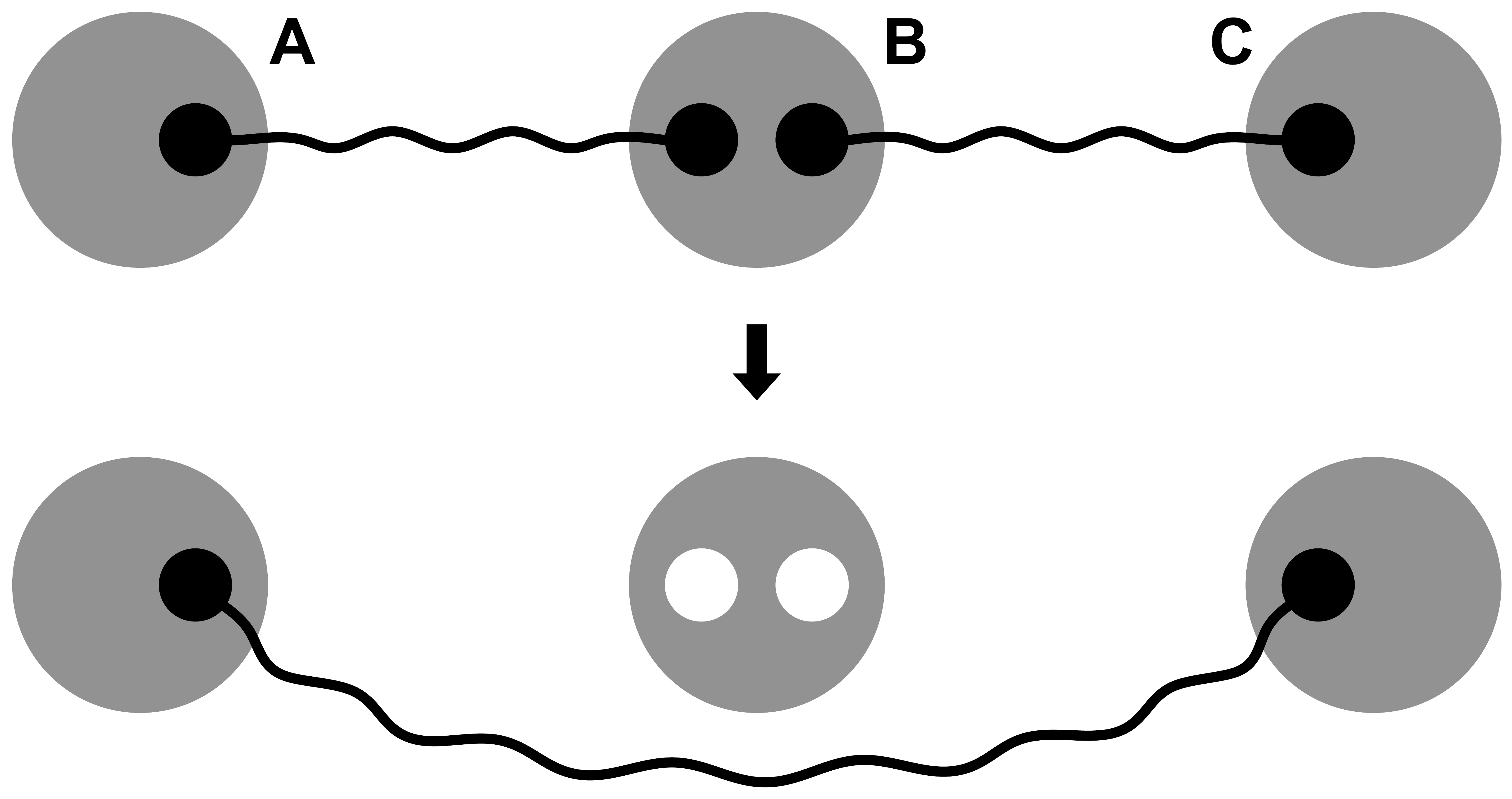}
	\centering
	\caption{\textbf{Entanglement swap.}
	When node B performs a swap, an entangled link between nodes A and B and an entangled link between nodes B and C are consumed to produce a single entangled link between A and C. This operation is essential for the generation of long-distance entanglement.}
	\label{fig.swap}
\end{figure}

\ssec{Entanglement swap.} Two neighboring entangled links can be fused into a longer-distance entangled link via entanglement swapping. Consider a situation where node B shares an entangled link with node A, and another link with node C (see Figure~\ref{fig.swap}). Then, B can perform an entanglement swap to produce an entangled link between A and C while consuming both initial links \cite{Zukowski1993, Duan2001, Sangouard2011}. We refer to the link generated in a swap operation as a swapped link.
This operation is also probabilistic: a new link is produced with probability $p_\text{s}$, and no link is produced (but both input links are still consumed) with probability $1-p_\text{s}$.

The generation of an entangled link between two end nodes without intermediate repeaters is limited by the distance between the end nodes \cite{Munro2015} -- e.g., the noise affecting a photon sent over an optical fiber grows exponentially with the length of the fiber \cite{Briegel1998}.
Therefore, a repeater chain that makes use of entanglement swapping is needed to generate end-to-end entanglement over long distances.
\\

\ssec{Cutoffs}. The fidelity of a quantum state decreases over time due to couplings to the environment \cite{Dur2005,Chirolli2008}.
These decoherence processes can be captured using a white noise model in which a depolarizing channel is applied to the entangled state at every instant. As a result, the fidelity of a Werner state at time $t$, $F(t)$, is given by
\begin{equation}\label{eq.ftT}
    F(t) = \frac{1}{4} + \Big( F(t-\Delta t) - \frac{1}{4}\Big) \mathrm{e}^{-\Delta t/\tau},
\end{equation}
where $\Delta t$ is an arbitrary interval of time and $\tau$ is a parameter that characterizes the exponential decay in fidelity of the whole entangled state due to the qubits being stored in noisy memories.
This parameter depends on the physical realization of the qubit. (\ref{eq.ftT}) is derived in Appendix~\ref{app.decoherence}.

In general, quantum network applications require quantum states with fidelity above some threshold value $F_\text{min}$.
A common solution is to impose a cutoff time $\tc$ on the entangled links: all entangled links used to generate the final end-to-end link must be generated within a time window of size $\tc$ \cite{Rozpedek2018}.
Imposing memory cutoffs requires keeping track of the time passed since the creation of each entangled link. We call this time the age of the link.
A link is discarded whenever it gets older than $\tc$.
Moreover, we assume that an entangled link generated as a result of entanglement swapping assumes the age of the oldest link that was involved in the swapping operation.
Another valid approach to calculate the age of a swapped link would be to re-compute the age based on the post-swap fidelity, although this would lead to a more complicated formulation to ensure that all the links that were used to produce a swapped link were generated within the time window of size $t_\text{cut}$.
To produce end-to-end links with fidelity above $F_\text{min}$ on a repeater chain that generates new links with fidelity $F_\text{new}$, it suffices to ensure that the sequence of events that produces the lowest end-to-end fidelity satisfies this requirement.
In Appendix~\ref{app.cutoff}, we show that such a sequence of events corresponds to all links being simultaneously generated in the first attempt and all the entanglement swaps being performed at the end of the $t_\text{cut}$ interval.
Analyzing such a sequence of events leads to the following condition for the cutoff time:
\begin{equation}\label{eq.cutoffcondition}
    \tc \leq -\tau \ln\Bigg(\frac{3}{4F_\text{new}-1} \Big( \frac{4F_\text{min}-1}{3} \Big)^{\frac{1}{n-1}} \Bigg),
\end{equation}
where $n$ is the number of nodes.
For a full derivation of the previous condition, see Appendix~\ref{app.cutoff}.
\\

In this paper, we consider quantum networks that operate with a limited number of qubits. Specifically, we use the following additional assumptions:
\begin{enumerate}[label=(\roman*)]
	\item The chain is \textbf{homogeneous}, i.e., the hardware is identical in all nodes. This means that all pairs of neighbors generate links with the same success probability $p$ and fidelity $F_\text{new}$, all swaps succeed with probability $p_\text{s}$, all states decohere according to some coherence time $\tau$, and all nodes apply the same cutoff time $t_\text{cut}$. This assumption may not hold for some long-distance quantum networks where each node is implemented using a different technology, but may be directly applicable to, e.g., small metropolitan-scale networks.
	\item We assume that each node has only \textbf{two storage qubits}, each of which is used to generate entanglement with one side of the chain. Each end node has a single storage qubit.
	This assumption is in line with the expectations for early quantum networks, in which nodes are likely to have a number of storage qubits on the order of the unit (e.g., in \cite{Pompili2021} the authors realized the first three-node quantum network using NV centers, each with a single storage qubit).
    	\item We also assume that classical communication between nodes is instantaneous. This means that every node has \textbf{global knowledge} of the state of the repeater chain in real time. In general, this is not a realistic assumption. However, given that classical communication delays decrease the performance of the network, our results constitute a lower bound on the expected delivery time of real setups and can be used as a benchmark.
	\item Time is discretized into non-overlapping \textbf{time slots}. During one time slot: ($i$) first, each pair of neighboring nodes attempts entanglement generation if they have free qubits; ($ii$) second, some time is allocated for the nodes to attempt entanglement swaps; and ($iii$) lastly nodes discard any entangled link that existed for longer than $\tc$ time slots.
	To decide if they want to perform a swap in the second part of the time step, nodes can take into account the state of the whole chain, including the results from entanglement generation within the same time slot, since classical communication is instantaneous.
	The unit of time used in this paper is the duration of a time slot, unless otherwise specified.
\end{enumerate}

A repeater chain under the previous assumptions is characterized by four parameters:
\begin{itemize}
	\item[·] $n$: number of nodes in the chain, including end nodes.
	\item[·] $p$: probability of successful entanglement generation.
	\item[·] $p_\text{s}$: probability of successful swap.
	\item[·] $\tc$: cutoff time.
	Note that $F_\text{new}$, $F_\text{min}$, and $\tau$ are used to determine a proper value of cutoff time (see condition (\ref{eq.cutoffcondition})), but they are not needed after that.
\end{itemize}

In an experimental setup, the value of $p$ is determined by the inter-node distance and the type of hardware used, as quantum nodes can be realized using different technologies, such as NV centers \cite{Rozpedek2019, Bernien2013, Hensen2015, Humphreys2018, Pompili2021} and trapped ions \cite{Moehring2007,Slodicka2013}.
Linear optics setups generally perform swaps with probability $p_\text{s}=0.5$ \cite{Duan2001,Calsamiglia2001}, while other setups can perform deterministic swaps ($p_\text{s}=1$) at the cost of a slower speed of operation \cite{Pompili2021}.
The cutoff time $\tc$ can be chosen by the user, as long as condition (\ref{eq.cutoffcondition}) is satisfied. Note that (\ref{eq.cutoffcondition}) depends on $\tau$ (which depends on the hardware available), $F_\text{new}$ (which depends on the hardware and the choice of entanglement generation protocol), and $F_\text{min}$ (which is specified by the final application).
\\

\begin{figure}[t]
	\includegraphics[width=0.8\textwidth]{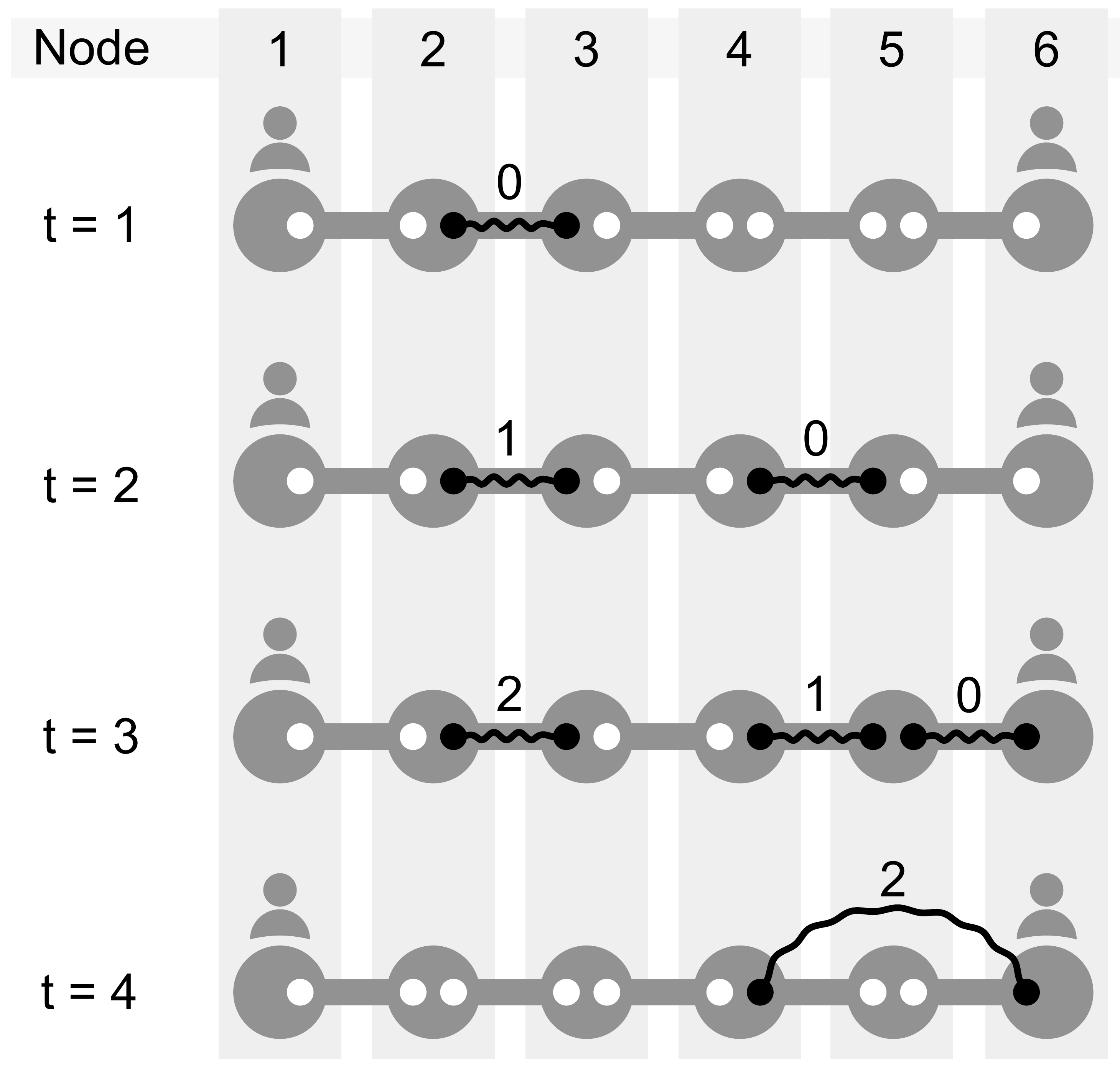}
	\centering
	\caption{\textbf{Example of entangled link dynamics in a repeater chain.}
	Each row represents the state of the chain at the end of time slot $t$.
	Entangled links are represented as black solid lines, with occupied qubits as black circles. The number above each entangled link is the age of the link. We assume cutoff $t_\text{cut}=3$.}
	\label{fig.chain_evolution}
\end{figure}

The state of the repeater chain at the end of each time slot can be described using the age of every entangled link.
In Figure~\ref{fig.chain_evolution} we show an example of the evolution of the state of a chain with cutoff $t_\text{cut}=3$, over four time slots:
\begin{itemize}
	\item In the first time slot ($t\in [0,1)$), all pairs of neighbors attempt entanglement generation, but it only succeeds between nodes two and three.
	No swaps can be performed, and the only link present is younger than the cutoff, so it is not discarded.
	\item In the second time slot ($t\in [1,2)$), the age of the link between nodes two and three increases by one. All pairs of neighbors (except nodes two and three) attempt entanglement generation, which succeeds between nodes four and five.
	\item In the third time slot ($t\in [2,3)$), the age of both existing links increases by one.
	All pairs of neighbors (except nodes two and three and nodes four and five) attempt entanglement generation, and only nodes five and six succeed.
	A swap can be performed at node five but they decide to wait.
	\item In the fourth time slot ($t\in [3,4)$), the age of every existing link increases by one.
	Nodes one and two and nodes three and four attempt entanglement generation but none of the pairs succeeds.
	A swap is successfully performed at node five, and a new link between nodes four and six is generated. This new link assumes the age of the oldest link involved in the swap operation.
	Lastly, the entangled link between nodes two and three is discarded, as its age reached the cutoff time.
\end{itemize}

\subsection{Optimal entanglement distribution policies}
As described above, nodes always attempt entanglement generation if there are qubits available.
Cutoffs are always applied whenever an entangled state becomes too old.
Since nodes are free to attempt swaps as soon as entangled links are available or some time later, they must agree on an entanglement distribution policy: a set of rules that indicate when to perform a swap.
An optimal policy minimizes the average time required to generate end-to-end entanglement when starting from any state (i.e., from any combination of existing links) and following said policy.
In particular, it minimizes the mean entanglement delivery time, which is the average time required to generate end-to-end entanglement when starting from the state with no entangled links.
We employ the mean entanglement delivery time as a performance metric.

In a global-knowledge policy, nodes have information about all the entangled links in the chain.
In a local-knowledge policy, the nodes only need to know the state of the qubits they hold.
An example of local policy is the swap-asap policy, in which each node performs a swap as soon as both entangled links are available.

We model the evolution of the state of the repeater chain as an MDP.
We then formulate the Bellman equations \cite{Sutton2018} and solve them using value iteration and policy iteration to find global-knowledge optimal policies.
More details and formal definitions are provided in Section \ref{sec.methods}.

\begin{figure}[t]
	\includegraphics[width=0.9\textwidth]{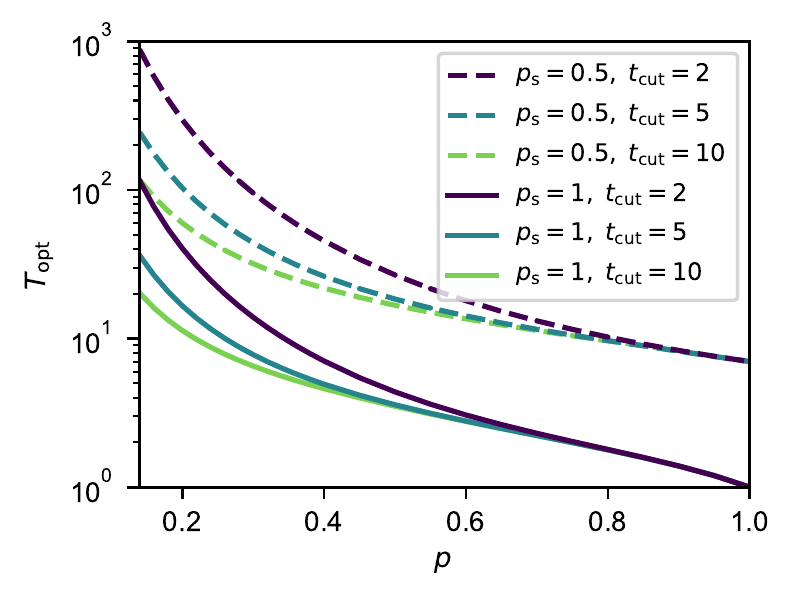}
\centering
\caption{\textbf{The expected delivery time increases with lower $p$, $p_\mathrm{s}$, and $t_\mathrm{cut}$.}
	Expected delivery time of an optimal policy, $T_\mathrm{opt}$, versus $p$ in a five-node chain, for different values of cutoff ($t_\text{cut}=2,5,10$). Solid lines correspond to deterministic swaps ($p_\text{s}=1$) and dashed lines correspond to probabilistic swaps with $p_\text{s}=0.5$.}
	\label{fig.delivery-time-n5}
\end{figure}

Let us now describe the relation between the expected delivery time of an optimal policy, $T_\mathrm{opt}$, and the variables of the system ($n$, $p$, $p_\mathrm{s}$, and $t_\text{cut}$).
Repeater chains with a larger number of nodes $n$ yield a larger $T_\mathrm{opt}$, since more entangled links need to be generated probabilistically.
When $p$ is small, more entanglement generation attempts are required to succeed, yielding a larger $T_\mathrm{opt}$.
Decreasing $p_\text{s}$ also increases $T_\mathrm{opt}$, since more attempts at entanglement swapping are required on average.
When $\tc$ is small, all entangled states must be generated within a small time window and therefore $T_\mathrm{opt}$ is also larger.
Figure~\ref{fig.delivery-time-n5} shows the expected delivery time of an optimal policy in a five-node chain.
Interestingly, $p_\text{s}$ has a much stronger influence on $T_\mathrm{opt}$ than $p$ and $\tc$: decreasing $p_\text{s}$ from 1 to 0.5 in a five-node chain translates into an increase in $T_\mathrm{opt}$ of an order of magnitude.
Similar behavior is observed for other values of $n$, as shown in Appendix~\ref{app.optimal}.

\begin{figure}[h]
	\includegraphics[width=\textwidth]{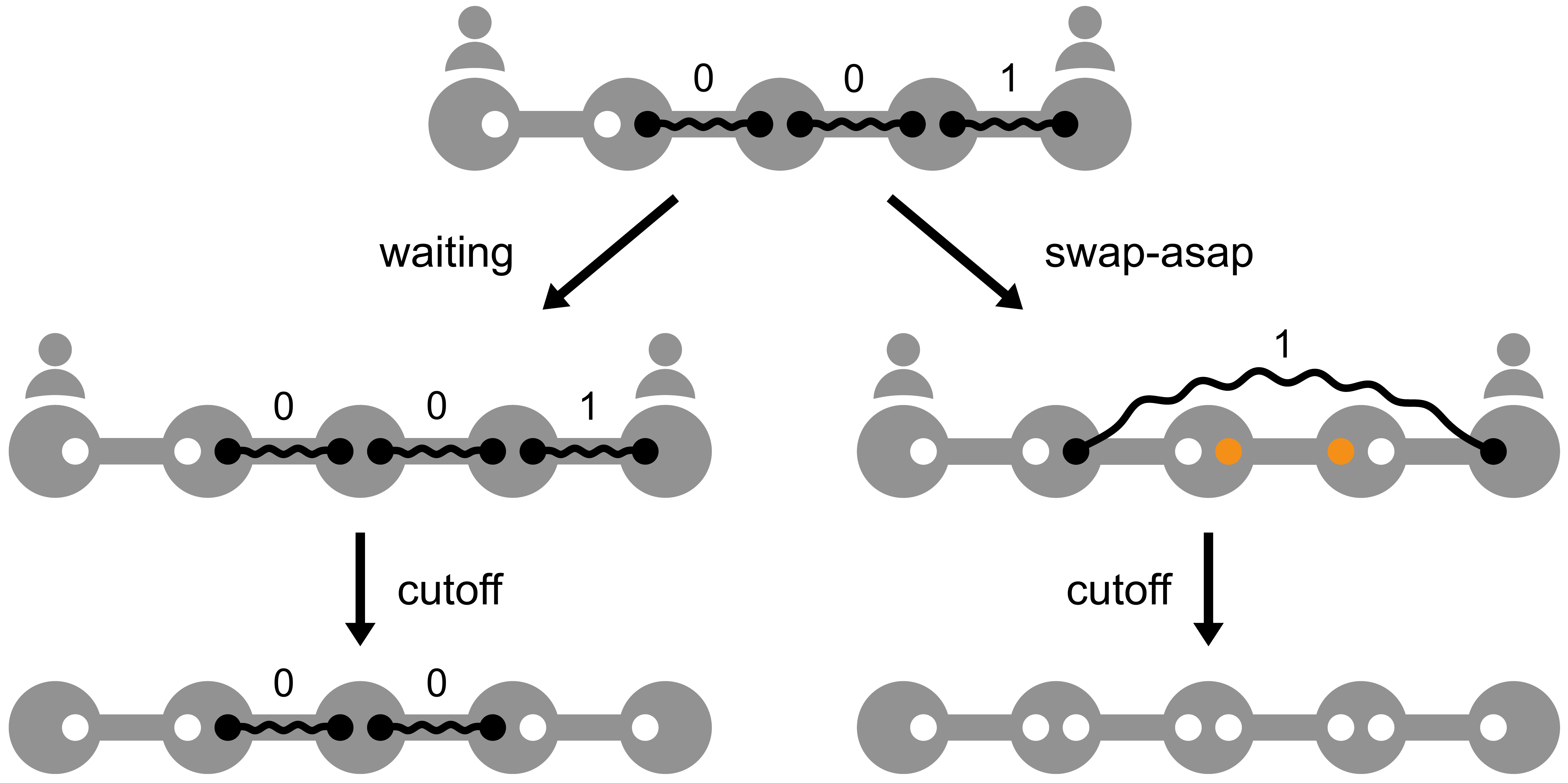}
\centering
\caption{\textbf{Swap-asap policies free up qubits, but swapped links expire earlier.}
Evolution of an example state when following a waiting policy versus the swap-asap policy during a single time slot.
Entangled links are represented as solid black lines, with occupied qubits in black and free qubits in white.
A waiting policy decides to not perform any swap, while the swap-asap policy decides to swap all three links.
The swap frees up qubits (marked in orange) that can be used to resume entanglement generation either if the swap is successful, as in the picture, or not.
After performing swaps, a cutoff $\tc=1$ is applied and links with age 1 are removed, causing the swapped link to expire.}
	\label{fig.chain_example}
\end{figure}

To evaluate the advantages of an optimal policy, we use the swap-asap policy as a baseline.
Early swaps can provide an advantage in terms of delivery time, since swapping earlier can free up qubits that can be used to generate backup entangled links, as displayed in the first transition in Figure~\ref{fig.chain_example}.
However, the age of a swapped link may reach the cutoff time earlier than one of the input links consumed in the swap, as the swapped link assumes the age of the oldest input link. Following the example in Figure~\ref{fig.chain_example} and assuming $\tc=1$, if no swaps are performed, the links between nodes two and three and between three and four will exist for one more time slot, while the link between nodes four and five will be removed immediately since it reached the cutoff time. If both swaps are performed, the swapped link between nodes two and five will be removed immediately since it reached the cutoff time.
Since we have arguments in favor of and against swapping early, it is not trivial to determine the scenarios in which the swap-asap policy is close to optimal. Next, we compare the expected delivery times of an optimal global-knowledge policy and the swap-asap policy.

\begin{figure}[t]
	\includegraphics[width=0.8\textwidth]{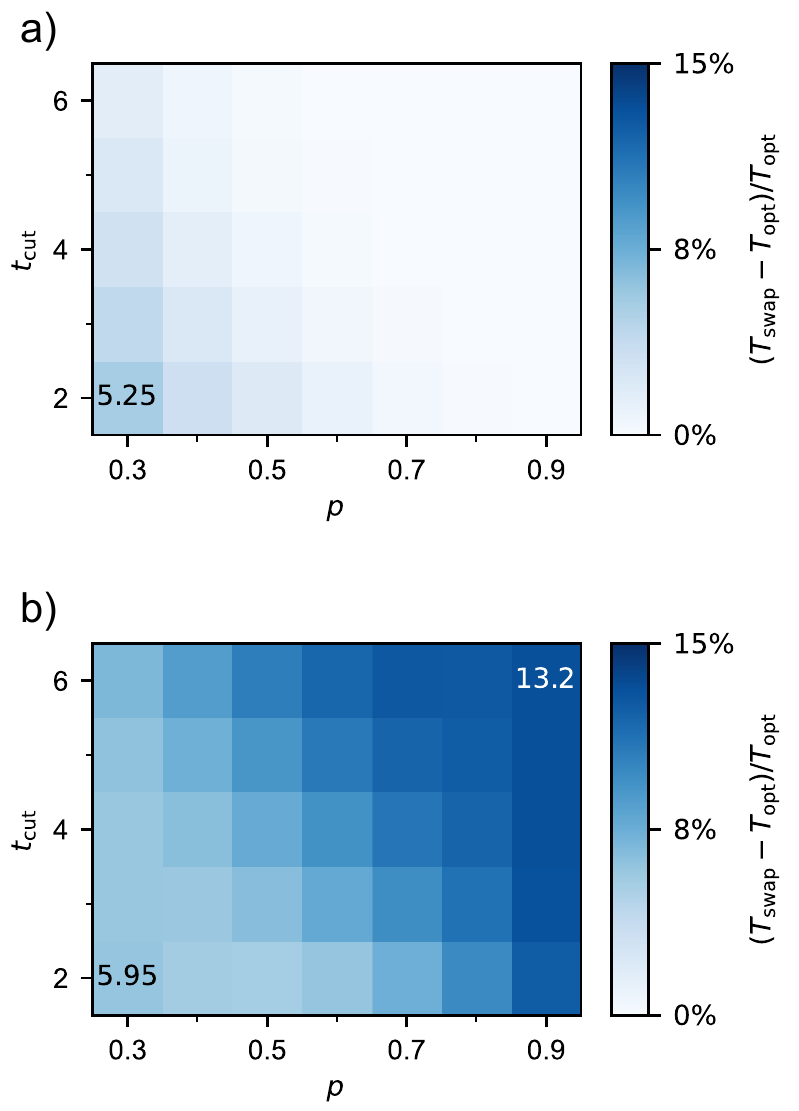}
     \centering
     	\caption{\textbf{In a five-node chain, an optimal policy performs increasingly better than swap-asap for lower/higher values of $p$ and $t_\mathrm{cut}$ when swaps are deterministic/probabilistic.}
	Relative difference between the expected delivery times of an optimal policy, $T_\mathrm{opt}$, and the swap-asap policy, $T_\mathrm{swap}$, in a five-node chain, for different values of $p$ and $\tc$.
	(a) Deterministic swaps ($p_\mathrm{s}=1$).
	(b) Probabilistic swaps ($p_\mathrm{s}=0.5$).
	}
	\label{fig.advantage5}
\end{figure}

Figure~\ref{fig.advantage5} shows the relative difference between the expected delivery times of an optimal global-knowledge policy, $T_\text{opt}$, and that of the swap-asap policy, $T_\text{swap}$, in a five-node chain.
Increasing values of $(T_\text{swap}-T_\text{opt})/T_\text{opt}$ mean that the optimal policy is increasingly faster on average.
Note that we restrict our analysis to the parameter regime $p\geq0.3$ and $2\leq\tc\leq6$ due to the very large computational cost of calculating the solution for smaller $p$ and larger $\tc$ (for more details, see Section \ref{sec.methods}).
Let us first focus on deterministic swaps (Figure~\ref{fig.advantage5}a).
The advantage provided by an optimal policy increases for decreasing $p$.
When $p$ is small, links are more valuable since they are harder to generate. Therefore, it is convenient to avoid early swaps, as they effectively increase the ages of the links involved and make them expire earlier.
When $\tc$ is small, a similar effect happens: all entangled links must be generated within a small time window and early swaps can make them expire too soon. For larger $\tc$, increasing the age of a link does not have a strong impact on the delivery time, since the time window is larger. Therefore, an optimal policy is increasingly better than swap-asap for decreasing $\tc$.
The maximum difference between expected delivery times in the parameter region explored is 5.25\%.

\begin{figure}[t]
	\includegraphics[width=\textwidth]{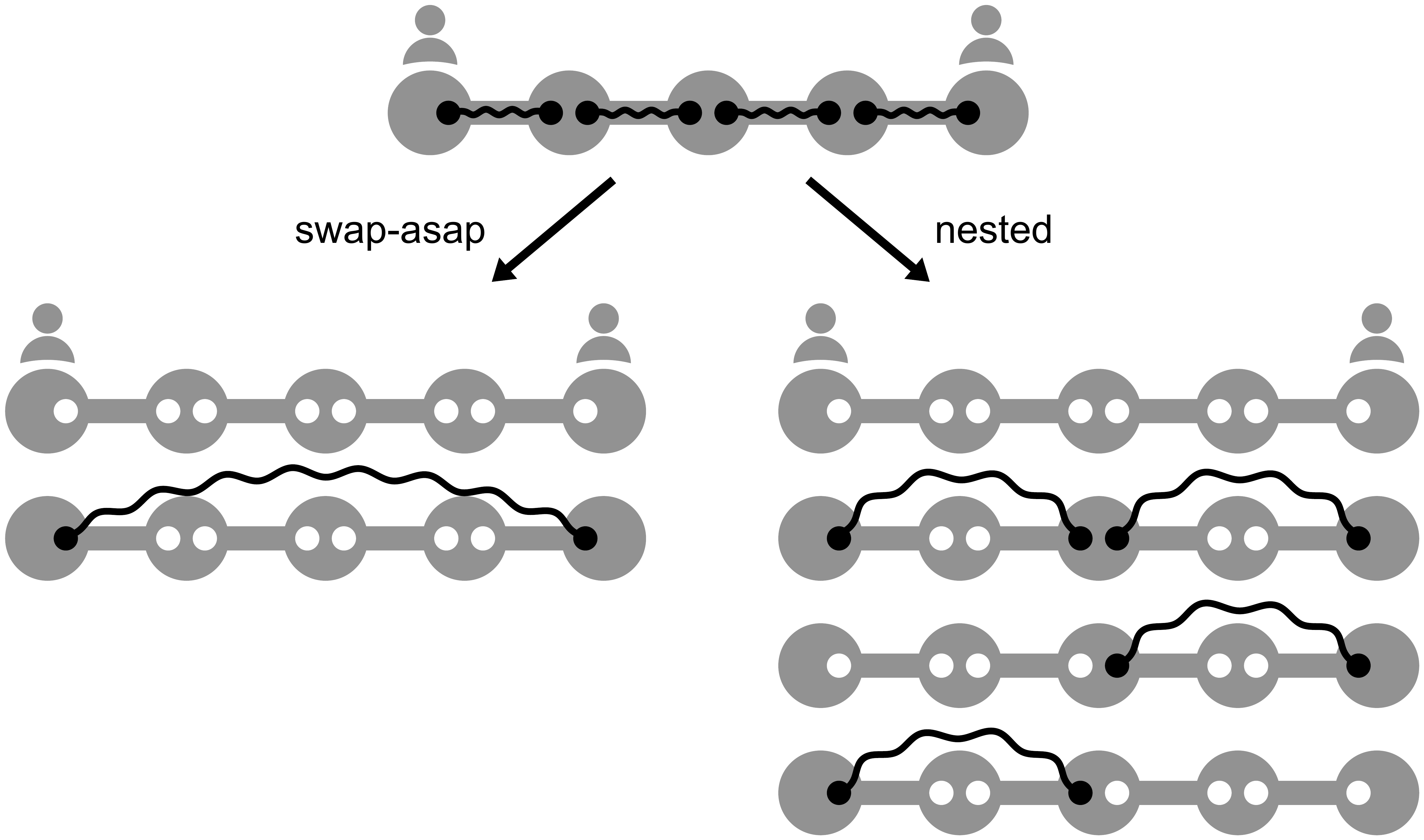}
	\centering
	\caption{\textbf{All possible transitions after performing a swap-asap action or a nested action in a full state, depending on which swaps succeed.}
	In full states, every pair of neighbors shares an entangled link (solid black lines, with occupied qubits in black and free qubits in white).
	The swap-asap policy decides to swap all links, while the nested approach consists in swapping only at nodes 2 and 4.
	When swaps are probabilistic, the nested approach is generally optimal in terms of expected delivery time.}
	\label{fig.swapasap-vs-doubling}
\end{figure}

Interestingly, probabilistic swaps (Figure~\ref{fig.advantage5}b) yield an opposite behavior in the parameter region explored: optimal policies are increasingly better than swap-asap for increasing $p$ and $\tc$ (except when $p\leq0.4$ and $\tc\leq3$), and the relative difference in expected delivery time can be as large as 13.2\% (achieved in a five-node chain with $p=0.9$ and $\tc=6$).
One reason for this may be the action that each policy decides to perform when the repeater chain is in a full state, which is a situation where each pair of neighboring nodes shares an entangled link (see state at the top of Figure~\ref{fig.swapasap-vs-doubling}).
When swaps are deterministic, the optimal policy chooses to swap all links in a full state, since end-to-end entanglement will always be achieved.
However, when swaps are probabilistic, an optimal policy generally chooses to perform two separate swaps (see Figure~\ref{fig.swapasap-vs-doubling}), similar to the nested purification scheme proposed in \cite{Briegel1998}.
As an example, for $n=5$, $p=0.9$, $\tc=2$, and $p_\text{s}=0.5$, the swap-asap policy yields an expected delivery time of $T=9.35$.
If, in full states, the swap at the third node is withheld, $T$ drops to 8.34.
The swap-asap policy is on average slower than this modified policy by 12.1\%.
The action chosen in full states has a stronger influence on $T$ for increasing $p$. This is because full states are more frequent for large $p$: whenever a swap fails, a full state is soon recovered, since new entangled states are generated with high probability.
As a consequence, an optimal policy is increasingly better than swap-asap for higher $p$ when swaps are probabilistic.
A similar effect happens for large $\tc$.
Note however that the effect of the action chosen in full states is practically irrelevant in four-node chains (see Appendix~\ref{app.optimal}).
Note also that the advantage of an optimal policy in terms of delivery time is not always monotonic in $p$ and $t_\text{cut}$ (see Appendix~\ref{app.optimal}).

\begin{figure}[t]
	\includegraphics[width=0.8\textwidth]{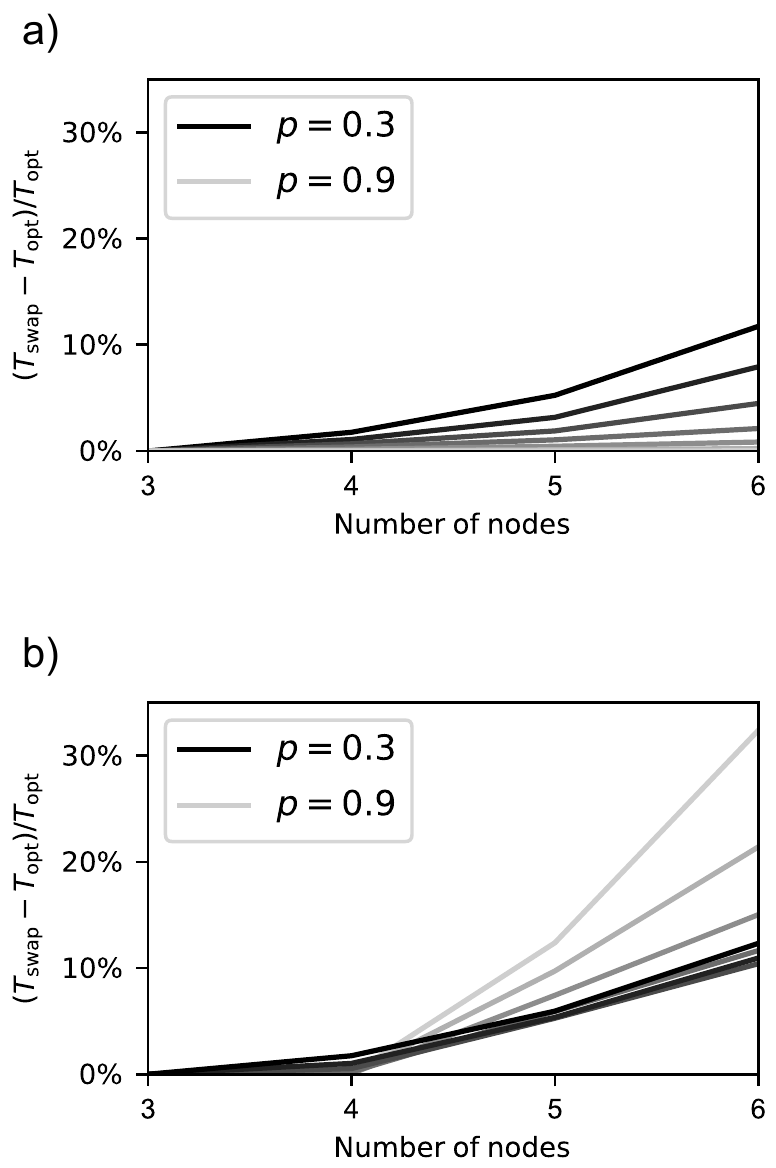}
	\centering
	\caption{\textbf{An optimal policy performs increasingly better than swap-asap in longer chains.}
	Relative difference between the expected delivery times of an optimal policy, $T_\mathrm{opt}$, and the swap-asap policy, $T_\mathrm{swap}$, for $\tc=2$ and different values of $p$, as a function of the number of nodes $n$.
	Black lines correspond to $p=0.3$, and the value of $p$ increases in steps of $0.1$ with increasing line transparency up to $p=0.9$.
	(a) Deterministic swaps ($p_\mathrm{s}=1$).
	(b) Probabilistic swaps ($p_\mathrm{s}=0.5$).}
	\label{fig.advantage-n}
\end{figure}

Optimal policies are also increasingly faster than swap-asap for increasing $n$, as shown in Figure~\ref{fig.advantage-n}.
For example, for $p=0.3$, $p_\text{s}=0.5$, and $t_\text{cut}=2$, the relative difference in expected delivery time is 1.7\%, 5.9\%, and 12.3\%, for $n=4$, 5, and 6, respectively. 
This is in line with the fact that, when the number of nodes grows, there are increasingly more states in which the optimal action to perform is a strict subset of all possible swaps, as shown in Appendix~\ref{app.swapasap-states}.
Note that, in three- and four-node chains, the relative difference in expected delivery time is generally below 1\%.

\section{Discussion}\label{sec.discussion}
Our work sheds light on how to distribute entanglement in quantum networks using a chain of intermediate repeaters with pre-configured cutoffs.
We have shown that optimal global-knowledge policies can significantly outperform other policies, depending on the properties of the network. In particular, we have found and explained non-trivial examples in which performing swaps as soon as possible is far from optimal.
We have also contributed a simple methodology to calculate optimal policies in repeater chains with cutoffs that can be extended to more realistic scenarios, e.g., asymmetric repeater chains, by modifying the transition probabilities of the MDP.

In this work, we have assumed that classical communication is instantaneous.
Hence, our optimal policies may become sub-optimal in setups with non-negligible communication times, where decisions must be made using local information only.
Nevertheless, our optimal policies still constitute a best-case policy against which to benchmark.

Note also that we have restricted our analysis to repeater chains with less than seven nodes. This is due to the exponentially large computational cost of solving the MDP for larger chains (see Appendix~\ref{app.dp} for further details).
However, each entanglement swap decreases the fidelity of the entangled links.
Hence, a large number of swaps limits the maximum end-to-end fidelity achievable, making chains with a very large number of nodes impractical.
Therefore, we consider the analysis of short chains to be more relevant.

An interesting extension of this work would be to explore different cutoff policies. For example, one could allow the nodes to decide when to discard entangled links, or one could optimize simultaneously over the cutoff and the swapping policy. This may lead to improved optimal policies.

As a final remark, note that we have employed the expected delivery time as the single performance metric. In some cases, the expected value and the variance of the delivery time distribution are within the same order of magnitude (some examples are shown in Appendix~\ref{app.delivery-time-distribution}). Therefore, an interesting follow-up analysis would be to study the delivery time probability distribution instead of only the expected value.
Additionally, we put fidelity aside by only requiring an end-to-end fidelity larger than some threshold value, via a constraint on the cutoff time.
This constraint can be lifted to optimize the fidelity instead of the expected delivery time, or to formulate a multi-objective optimization problem to maximize fidelity while minimizing delivery time.

\section{Methods}\label{sec.methods}
We have formulated the problem of finding optimal entanglement distribution policies as an MDP where each state is a combination of existing entangled links and link ages.
Let $\boldsymbol{s}$ be the state of the repeater chain at the beginning of a time slot.
As previously explained, $\boldsymbol{s}$ can be described using the age of every entangled link. Mathematically, this means that $\boldsymbol{s}$ can be represented as a vector of size $n\choose 2$:
$$ \boldsymbol{s} = [ g_1^2, g_1^3, \dots, g_1^n;\; g_2^3, \dots, g_2^n;\; \dots;\; g_{n-1}^n ],$$
where $g_i^j$ is the age of the entangled link between nodes $i$ and $j$ (if nodes $i$ and $j$ do not share an entangled link, then $g_i^j=-1$).
In each time slot, the nodes must choose and perform an action $a$. Mathematically, $a$ is a set containing the indices of the nodes that must perform swaps (if no swaps are performed, $a=\emptyset$).

The state of the chain at the end of the time slot is $\boldsymbol{s}'$.
Since entanglement generation and swaps are probabilistic, the transition from $\boldsymbol{s}$ to $\boldsymbol{s}'$ after performing $a$  happens with some transition probability $P(\boldsymbol{s}'|\boldsymbol{s},a)$.
A policy is a function $\pi$ that indicates the action that must be performed at each state, i.e.,
$$\pi \colon \boldsymbol{s} \in \mathcal{S} \to \pi(\boldsymbol{s})\in \mathcal{A} \, ,$$
where $\mathcal{S}$ is the state space and $\mathcal{A}$ is the action space.
W.l.o.g., we only consider deterministic policies, otherwise a policy would be a probability distribution instead of a function (see Appendix~\ref{app.hitting_time} for further details).

Let us define $\boldsymbol{s}_0$ as the state where no links are present and $\mathcal{S}_\text{end}$ as the set of states with end-to-end entanglement, also called absorbing states.
In general, the starting state is $\boldsymbol{s}_0$, and the goal of the repeater chain is to transition to a state in $\mathcal{S}_\text{end}$ in the fewest number of steps. When a state in $\mathcal{S}_\text{end}$ is reached, the process stops.
Let us define the expected delivery time from state $\boldsymbol{s}$ when following policy $\pi$, $T_\pi(\boldsymbol{s})$, as the expected number of steps required to reach an absorbing state when starting from state $\boldsymbol{s}$.
The expected delivery time is also called hitting time in the context of Markov chains (see Chapter 9 from \cite{VanMieghem2014}).
A policy $\pi$ is better than or equal to a policy $\pi'$ if $T_\pi(\boldsymbol{s}) \leq T_{\pi'}(\boldsymbol{s})$, $\forall \boldsymbol{s}\in\mathcal{S}$.
An optimal policy $\pi^*$ is one that is better than or equal to all other policies. In other words, an optimal policy is one that minimizes the expected delivery time from all states.
One can show that there exists at least one optimal policy in an MDP with a finite and countable set of states (see Section 2.3 from \cite{Szepesvari2010}).
To find such an optimal policy, we employ the following set of equations, which are derived in Appendix~\ref{app.hitting_time}:
\begin{equation}\label{eq.T_s}
    T_\pi(\boldsymbol{s}) = 1 + \sum_{\boldsymbol{s}'\in\mathcal{S}} P(\boldsymbol{s}'|\boldsymbol{s},\pi) \cdot T_\pi(\boldsymbol{s}'), \; \forall \boldsymbol{s}\in\mathcal{S},
\end{equation}
where $\mathcal{S}$ is the state space and $P(\boldsymbol{s}'|\boldsymbol{s},\pi)$ is the probability of transition from state $\boldsymbol{s}$ to state $\boldsymbol{s}'$ when following policy $\pi$.
Equations (\ref{eq.T_s}) are a particular case of what is generally known in the literature as the Bellman equations.

An optimal policy can be found by minimizing $T_\pi(\boldsymbol{s})$, $\forall \boldsymbol{s}\in\mathcal{S}$, using (\ref{eq.T_s}).
To solve this optimization problem, we used value iteration and policy iteration, which are two different iterative methods whose solution converges to the optimal policy (both methods provided the same results).
For more details, see Appendix~\ref{app.dp}, and for a general reference on value and policy iteration, see Chapter~4 from \cite{Sutton2018}.

We provide an example of how to calculate the transition probabilities $P(\boldsymbol{s}'|\boldsymbol{s},\pi)$ analytically in Appendix~\ref{app.example}, although this is generally impractical, since the size of the state space grows at least exponentially with $n$ and polynomially with $\tc$ (as shown in Appendix~\ref{app.scaling}, $|\mathcal{S}| = \Omega\big((\tc)^{n-2}\big)$).
Lastly, in Appendix~\ref{app.environment} we discuss how to simplify the calculation of transition probabilities.

As a validation check, we also implemented a Monte Carlo simulation that can run our optimal policies, providing the same expected delivery time that we obtained from solving the MDP.

\section{Data availability}\label{sec.data_availability}
The data shown in this paper can be found at \cite{Inesta2022dataset}.

\section{Code availability}\label{sec.code_availability}
Our code can be found in the following GitHub repository: \href{https://github.com/AlvaroGI/optimal-homogeneous-chain}{https://github.com/AlvaroGI/optimal-homogeneous-chain}.

\section{Acknowledgments}
We thank Subhransu Maji, Guus Avis, and Bethany Davies for discussions and feedback.
\'{A}GI acknowledges financial support from the Netherlands Organisation for Scientific Research (NWO/OCW), as part of the Frontiers of Nanoscience program.
GV acknowledges financial support from the NWO ZK QSC Ada Lovelace Fellowship.
SW acknowledges support from an ERC Starting Grant.

\section{Author contributions}
\'{A}GI and GV defined the problem, the model, and the MDP formulation.
\'{A}GI and GV, with support from LS, coded iterative methods to solve the MDP.
\'{A}GI analyzed the results and was the main writer of this paper.
SW provided active feedback at every stage of the project.

\section{Competing interests}
The authors declare no competing interests.


%


\clearpage

\onecolumngrid
\appendix

\vspace{-20pt}
\section{Depolarization of Werner states}\label{app.decoherence}
In this Appendix we show that the fidelity of a Werner state in which each qubit independently experiences a depolarizing process evolves as
$$F(t) = \frac{1}{4} + \bigg(F(t-\Delta t)-\frac{1}{4}\bigg) \mathrm{e}^{-\frac{\Delta t}{\tau}},$$
where $t$ is the time, $\Delta t$ is an arbitrary interval of time, and $\tau$ is a parameter that characterizes the exponential decay in fidelity of the whole entangled state due to the qubits being stored in noisy memories.
Note that we assume independent noise on each qubit since, in our problem, they are stored in different nodes of the repeater chain.

The depolarizing channel \cite{Nielsen2002, Dur2005} is defined as
\begin{equation}
	\mathcal{E}_i: \;\; \rho_i \;\;\rightarrow\;\; p \rho_i + (1-p) \frac{\mathbb{I}_2}{2},
\end{equation}
where $\rho_i$ is a single-qubit state, $0 \leq p \leq 1$ (this $p$ is not to be confused with the entanglement generation probability used in the main text of this paper), and $\mathbb{I}_d$ is the $d$-dimensional identity.
Let us assume that each qubit independenlty experiences a depolarizing channel while stored in memory for a finite time $t_\mathrm{dep}$.
During an interval of time $t_\mathrm{dep}$, a Werner state $\rho$ with fidelity $F$ is therefore mapped to $(\mathcal{E}_1\otimes\mathcal{E}_2) (\rho)$, where $\mathcal{E}_i$ is a depolarizing channel acting on the $i$-th qubit.
Let us calculate this output state explicitly:
\begin{equation}\label{eq.Fbasecase}
	\begin{split}
		(\mathcal{E}_1\otimes\mathcal{E}_2) (\rho) =\;&p^2 \rho + p(1-p) \Tr_2(\rho) \otimes \frac{\mathbb{I}_2}{2}
					 + p(1-p) \frac{\mathbb{I}_2}{2} \otimes  \Tr_1(\rho)
					+ (1-p)^2 \frac{\mathbb{I}_4}{4}\\
			\stackrel{a}{=}\;&p^2 \rho + p(1-p) \frac{\mathbb{I}_2}{2} \otimes \frac{\mathbb{I}_2}{2}
					 + p(1-p) \frac{\mathbb{I}_2}{2} \otimes  \frac{\mathbb{I}_2}{2}
					+ (1-p)^2 \frac{\mathbb{I}_4}{4}\\
			=\;&p^2 \rho + \big(2p(1-p) + (1-p)^2\big) \frac{\mathbb{I}_4}{4}\\
			\stackrel{b}{=}\;&p^2 \frac{4F-1}{3} \ketbra{\phi^+} + p^2\frac{1-F}{3}\mathbb{I}_4
					 + \big(2p(1-p) +(1-p)^2\big) \frac{\mathbb{I}_4}{4}\\
			\stackrel{c}{=}\;& \frac{4F'-1}{3} \ketbra{\phi^+} + \frac{1-F'}{3}\mathbb{I}_4,
	\end{split}
\end{equation}
with the following steps:
\begin{enumerate}[label=\alph*.]
	\item We use the fact that the partial trace of a maximally entangled state is a maximally mixed state. As a consequence, $\Tr_i(\rho)=\frac{\mathbb{I}_2}{2}$, for any Werner state $\rho$.
	\item We use the definition of Werner state: $\rho = \frac{4F-1}{3}\ketbra{\phi^+} + \frac{1-F}{3} \mathbb{I}_4$.
	\item We define $F' = \frac{1}{4} + p^2(F-\frac{1}{4})$.
\end{enumerate}

The output state $(\mathcal{E}_1\otimes\mathcal{E}_2) (\rho)$ is a Werner state with fidelity $F'$.
Then, the application of $n$ successive transformations $\mathcal{E}_1\otimes\mathcal{E}_2$ produces a Werner state with fidelity
\begin{equation}\label{eq.Fn}
	F^{(n)} = \frac{1}{4} + p^{2n}\bigg(F-\frac{1}{4}\bigg).
\end{equation}
This can be shown by induction as follows. The base case is proven in (\ref{eq.Fbasecase}): $F^{(1)} = \frac{1}{4} + p^2(F-\frac{1}{4})$. Next, if we assume that (\ref{eq.Fn}) is true for $n=k$, we can show that it also holds for $n=k+1$:
\begin{equation*}
\begin{split}
	F^{(k+1)} &= \frac{1}{4} + p^{2}\bigg(F^{(k)}-\frac{1}{4}\bigg) 
	= \frac{1}{4} + p^{2}\bigg(\frac{1}{4} + p^{2k}\Big(F-\frac{1}{4}\Big)-\frac{1}{4}\bigg) 
	= \frac{1}{4} + p^{2(k+1)}\bigg(F-\frac{1}{4}\bigg),
\end{split}
\end{equation*}
where we have used (\ref{eq.Fbasecase}) in the first step.

The total time required for these operations is $\Delta t=nt_\mathrm{dep}$. Therefore, if the fidelity of the state at time $t-\Delta t$ was $F(t-\Delta t)$, the fidelity at $t$ is given by
\begin{equation}
	F(t) = \frac{1}{4} + p^{2\Delta t/t_\mathrm{dep}}\bigg(F(t-\Delta t)-\frac{1}{4}\bigg).
\end{equation}

Finally, we map $p\in[0,1]$ to a new parameter $\tau\in[0,+\infty)$ as $p^2 \equiv \mathrm{e}^{-t_\mathrm{dep}/\tau}$.
Then, we obtain
\begin{equation}
	F(t) = \frac{1}{4} + \bigg(F(t-\Delta t)-\frac{1}{4}\bigg) \mathrm{e}^{-\frac{\Delta t}{\tau}}.
\end{equation}

\vspace{20pt}

\section{Relation between cutoff and threshold fidelity}\label{app.cutoff}
In the design of a quantum repeater chain, we must select a cutoff time $\tc$ such that the fidelity of any end-to-end entangled link is larger than some threshold $F_\text{min}$.
We show that this requirement is always satisfied when the cutoff time meets the following condition:
\begin{equation}
    \tc \leq -\tau \ln\Bigg(\frac{3}{4F_\text{new}-1} \Big( \frac{4F_\text{min}-1}{3} \Big)^{\frac{1}{n-1}} \Bigg),
\end{equation}
where $\tau$ is a parameter that characterizes the exponential decay in fidelity of the whole entangled state due to the qubits being stored in noisy memories, $F_\text{new}$ is the fidelity of newly generated entangled links, $F_\text{min}$ is the minimum desired end-to-end fidelity, and $n$ is the number of nodes in the chain.

First, we analyze the impact of a late entanglement swap on the fidelity of the output state.
As shown in Appendix~\ref{app.decoherence}, the fidelity of a Werner state that experiences depolarizing noise independently on each qubit decays as
\begin{equation}\label{eq.Fdecay}
    F(t) = \frac{1}{4} + \Big( F(t-\Delta t) - \frac{1}{4}\Big) \mathrm{e}^{-\Delta t/\tau},
\end{equation}
over an interval of time time $\Delta t$.
When two Werner states are used as input in an entanglement swap, the output state is a Werner state with fidelity
\begin{equation}\label{eq.swap}
    F_\text{swap}(F_1,F_2) = F_1 \cdot F_2 + \frac{(1-F_1)\cdot(1-F_2)}{3},
\end{equation}
where $F_1$ and $F_2$ are the fidelities of the input states \cite{Munro2015}.

Let us consider two Werner states with initial fidelities $F_1(t_0)$ and $F_2(t_0)$, respectively.
On the one hand, if we perform a swap and then wait for some time $t_\mathrm{wait}$, the final state is a Werner state with fidelity
\begin{equation}\label{eq.Fswapwait}
    F_\text{swap-wait}(t_0+t_\mathrm{wait}) = \frac{1}{4} + \bigg( F_1(t_0) F_2(t_0) - \frac{1}{4} 
    	+ \frac{\big(1-F_1(t_0)\big)\big(1-F_2(t_0)\big)}{3}\bigg) \mathrm{e}^{-t_\mathrm{wait}/\tau},
\end{equation}
which can be obtained by applying (\ref{eq.swap}) first and (\ref{eq.Fdecay}) next.
On the other hand, if we wait for some time $t_\mathrm{wait}$ and then perform the swap, we obtain a Werner state with fidelity
\begin{equation}\label{eq.Fwaitswap}
\begin{split}
    F_\text{wait-swap}(t_0+t_\mathrm{wait}) &= F_1(t_0+t_\mathrm{wait}) F_2(t_0+t_\mathrm{wait})
	+ \frac{\big(1-F_1(t_0+t_\mathrm{wait})\big)\big(1-F_2(t_0+t_\mathrm{wait})\big)}{3}\\
    &= \frac{1}{4} + \Bigg( F_1(t_0) F_2(t_0)  - \frac{1}{4}
	+ \frac{\big(1-F_1(t_0)\big)\big(1-F_2(t_0)\big)}{3} \Bigg) \mathrm{e}^{-2t_\mathrm{wait}/\tau},
\end{split}
\end{equation}
where we have used (\ref{eq.Fdecay}) in the second step and performed some basic algebra.
Note that the factor that multiplies the exponential in (\ref{eq.Fswapwait}) and (\ref{eq.Fwaitswap}) is nonnegative as long as $F_1(t_0), F_2(t_0)\geq\frac{1}{4}$ -- if the initial fidelity is $\frac{1}{4}$, the initial state is a maximally mixed state.

By comparing (\ref{eq.Fswapwait}) and (\ref{eq.Fwaitswap}), we find that an entangled link with larger fidelity is obtained if we first perform an entanglement swap and then wait for time $t_\mathrm{wait}$ rather than if we wait for time $t_\mathrm{wait}$ and then perform the swap, since
\begin{equation}\label{eq.swapwait}
    F_\text{swap-wait}(t_0+t_\mathrm{wait}) > F_\text{wait-swap}(t_0+t_\mathrm{wait}), \;\;\forall t_\mathrm{wait}>0.
\end{equation}

Let us now consider a sequence of $m$ entangled links that can be fused into a single long link after performing $m-1$ swaps.
Each of the initial links has fidelity $F_i$, $i=1,...,m$.
We want to calculate the final fidelity, assuming that all swaps are successful.
For this, it is convenient to define a Werner state in terms of the Werner parameter $x$:
\begin{equation}
    \rho = x\ketbra{\phi^+} + \frac{1-x}{4} \mathbb{I}_4,
\end{equation}
where $\ket{\phi^+}=\frac{\ket{00}+\ket{11}}{\sqrt{2}}$ is a Bell state, and $\mathbb{I}_d$ is the $d$-dimensional identity. The Werner parameter $x$ is defined in terms of the fidelity as $x=\frac{4F-1}{3}$.
Equation (\ref{eq.swap}) can be written in terms of the Werner parameter of each state:
\begin{equation}\label{eq.swapx}
    x_\text{swap}(x_1, x_2) = x_1 x_2,
\end{equation}
where $x_\text{swap}$ is the Werner parameter of the output state after swapping two Werner states with Werner parameters $x_1$ and $x_2$.
If we apply Equation (\ref{eq.swapx}) repeatedly to our sequence of $m$ entangled links, assuming that all swaps happen simultaneously (i.e., with no decoherence happening in between swaps), we obtain a final state with Werner parameter
\begin{equation}
	x_\text{final} = x_1 x_2 \dots x_m = \prod_{i=1}^m \frac{4F_i-1}{3}.
\end{equation}
Then, the final fidelity is given by
\begin{equation}\label{eq.Ffinal}
	F_\text{final} = \frac{3x_\text{final}+1}{4} = \frac{1}{4} + \frac{3}{4} \prod_{i=1}^m \frac{4F_i-1}{3}.
\end{equation}
A similar result was derived in \cite{Briegel1998}, although assuming $F_i = F$, $\forall i$.

We are now ready to find a relationship between the cutoff time and the minimum fidelity in an $n$-node quantum repeater chain with cutoff time $\tc$.
For this, we need to identify the sequence of events that produces the end-to-end link with the lowest fidelity.
First, note that all the entangled links that eventually form a single end-to-end link are created within a window of $\tc$ time slots.
According to (\ref{eq.swapwait}), delaying entanglement swaps has a negative impact on the final fidelity.
Therefore, the sequence of events that produces end-to-end entanglement with the lowest fidelity must be one where all swaps are performed at the end of the time window, i.e., all swaps are performed when the oldest link reaches the cutoff time, just before it expires. If any of those swaps were performed earlier, the final fidelity would be larger.
Such a sequence of events produces the lowest end-to-end fidelity when all the links are as old as possible, i.e., when their age is $t_\mathrm{cut}$. If any of the links were younger, the end-to-end fidelity would be larger.
Hence, the lowest end-to-end fidelity is achieved when all the links are generated simultaneously and all the swaps are performed when those links are  $t_\mathrm{cut}$ time slots old.
In this case, the fidelity of each link before swapping is given by (\ref{eq.Fdecay}):
\begin{equation}\label{eq.decohere_elem_link}
    F_\text{old} = \frac{1}{4}+\Big(F_\text{new}-\frac{1}{4}\Big)\mathrm{e}^{-\frac{\tc}{\tau}},
\end{equation}
where $F_\text{new}$ is the fidelity of newly generated elementary links.
The final fidelity after swapping all the links can be calculated using (\ref{eq.Ffinal}):
\begin{equation}
	F_\text{worst} = \frac{1}{4}\cdot\bigg[ 1 + \frac{(4F_\text{old}-1)^{n-1}}{3^{n-2}} \bigg].
\end{equation}

Finally, we impose that the worst-case end-to-end fidelity must be larger than the desired minimum fidelity $F_\text{min}$: $F_\text{worst} \geq F_\text{min}$.
Solving for $\tc$ yields an explicit condition for the cutoff time:
\begin{equation}\label{eq.tcutfinal}
    \tc \leq -\tau \ln\Bigg(\frac{3}{4F_\text{new}-1} \Big( \frac{4F_\text{min}-1}{3} \Big)^{\frac{1}{n-1}} \Bigg).
\end{equation}

When this condition is satisfied, every sequence of events will lead to a large enough fidelity. Consequently, any policy that we implement on the repeater chain will also deliver entanglement with a large enough fidelity.

\vspace{20pt}

\section{Further comments on the expected delivery time of optimal policies}\label{app.optimal}
Here we provide the expected delivery time of optimal policies in three- and four-node repeater chains. Then, we compare optimal policies to the swap-asap policy in a four-node chain. We also show that, in longer chains, the relative difference in expected delivery time is not always monotonic with the probability of successful entanglement generation $p$.

\begin{figure}[h]
\captionsetup[subfigure]{justification=centering}
     \centering
     \begin{subfigure}[b]{0.4\textwidth}
         \centering
         \includegraphics[width=\textwidth]{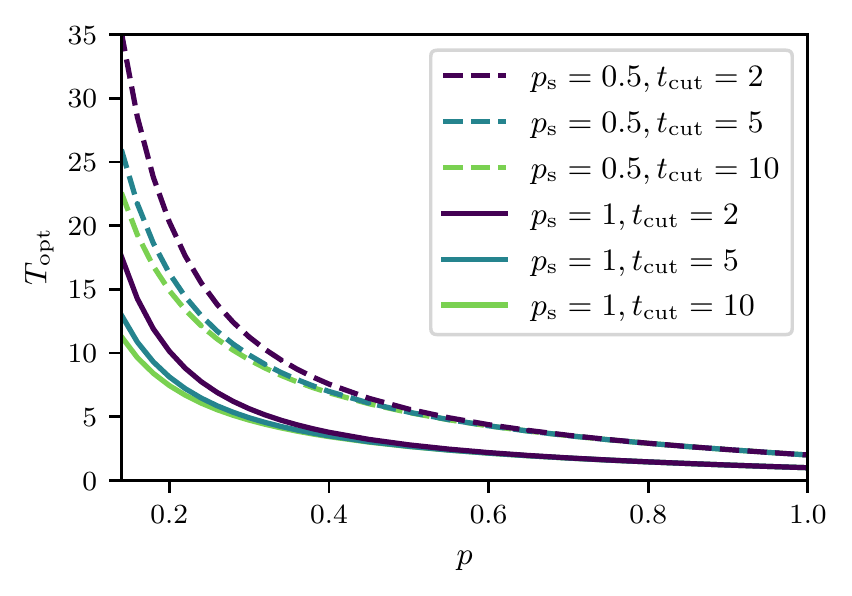}
         \caption{Three-node chain.}
         \label{fig.delivery-time-n3}
     \end{subfigure}
     \begin{subfigure}[b]{0.4\textwidth}
         \centering
         \includegraphics[width=\textwidth]{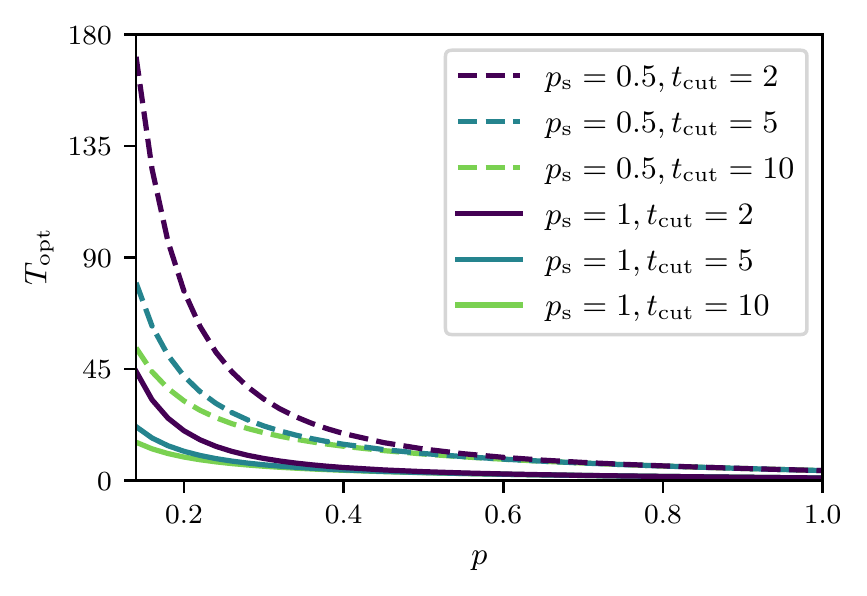}
         \caption{Four-node chain.}
         \label{fig.delivery-time-n4}
     \end{subfigure}
     
	\caption{The expected delivery time increases with lower $p$, $p_\mathrm{s}$, and $t_\mathrm{cut}$.
	Expected delivery time of an optimal policy, $T_\mathrm{opt}$, versus $p$ for (a) $n=3$ and (b) $n=4$ and different values of cutoff ($t_\text{cut}=2,5,10$). Solid lines correspond to deterministic swaps ($p_\text{s}=1$) and dashed lines correspond to probabilistic swaps with $p_\text{s}=0.5$.}
	\label{fig.delivery-time-n34}
\end{figure}

Figure~\ref{fig.delivery-time-n34} shows the expected delivery time of an optimal policy, $T_\text{opt}$, in three- and four-node chains, versus $p$ for different values of $p_\mathrm{s}$ and $\tc$.
The relation between $T_\text{opt}$ and the rest of the variables is similar to that of the five-node chain discussed in the main text.
When $p$ is small, more entanglement generation attempts are required to succeed, yielding a larger $T_\mathrm{opt}$.
Decreasing $p_\text{s}$ also increases $T_\mathrm{opt}$, since more attempts at entanglement swapping are required on average.
When $\tc$ is small, all entangled states must be generated within a small time window and therefore $T_\mathrm{opt}$ is also larger.

\begin{figure}[h]
\captionsetup[subfigure]{justification=centering}
     \centering
     \begin{subfigure}[b]{0.4\textwidth}
         \centering
         \includegraphics[width=\textwidth]{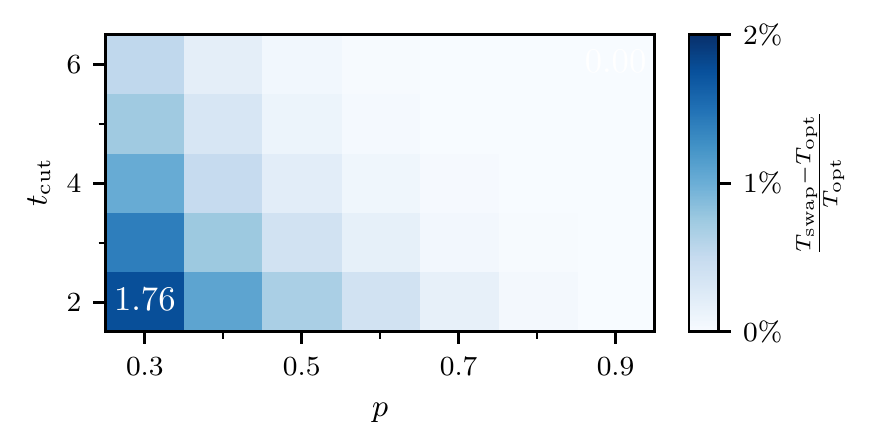}
         \caption{Deterministic swaps ($p_\mathrm{s}=1$).}
	\label{fig.advantage4-ps1}
     \end{subfigure}
     \begin{subfigure}[b]{0.4\textwidth}
         \centering
         \includegraphics[width=\textwidth]{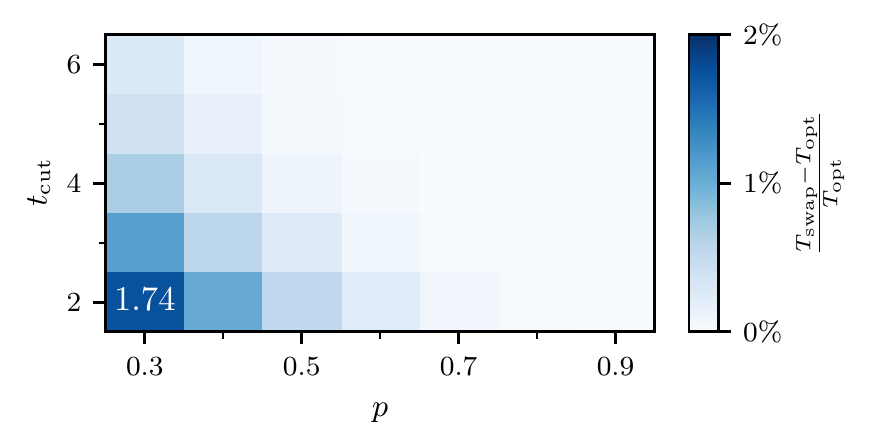}
         \caption{Probabilistic swaps ($p_\mathrm{s}=0.5$).}
	\label{fig.advantage4-ps05}
     \end{subfigure}
     
	\caption{The swap-asap policy is close to optimal in four-node chains.
	Relative difference between the expected delivery times of an optimal policy, $T_\mathrm{opt}$, and the swap-asap policy, $T_\mathrm{swap}$, in a four-node chain, for different values of $p$ and $\tc$.
	}
	\label{fig.advantage4}
\end{figure}

\begin{figure}[t]
\captionsetup[subfigure]{justification=centering}
     \centering
     \begin{subfigure}[b]{0.4\textwidth}
         \centering
         \includegraphics[width=\textwidth]{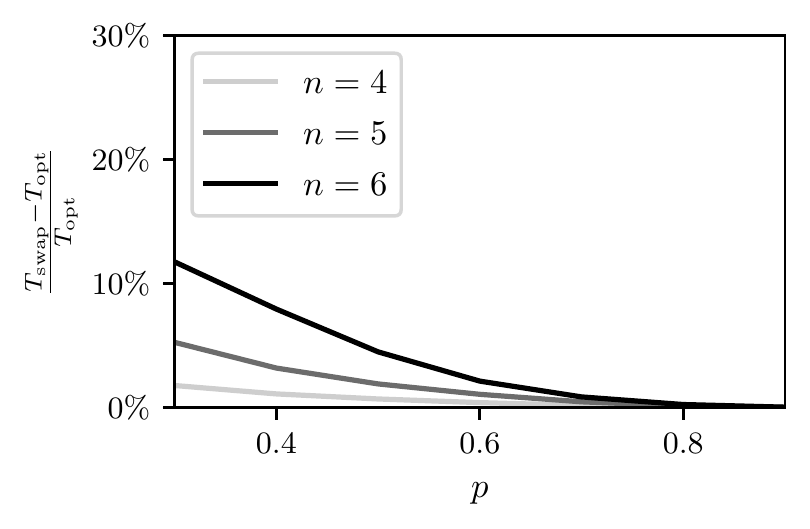}
         \caption{Deterministic swaps ($p_\mathrm{s}=1$).}
	\label{fig.advantage-p-different-n-ps1}
     \end{subfigure}
     \begin{subfigure}[b]{0.4\textwidth}
         \centering
         \includegraphics[width=\textwidth]{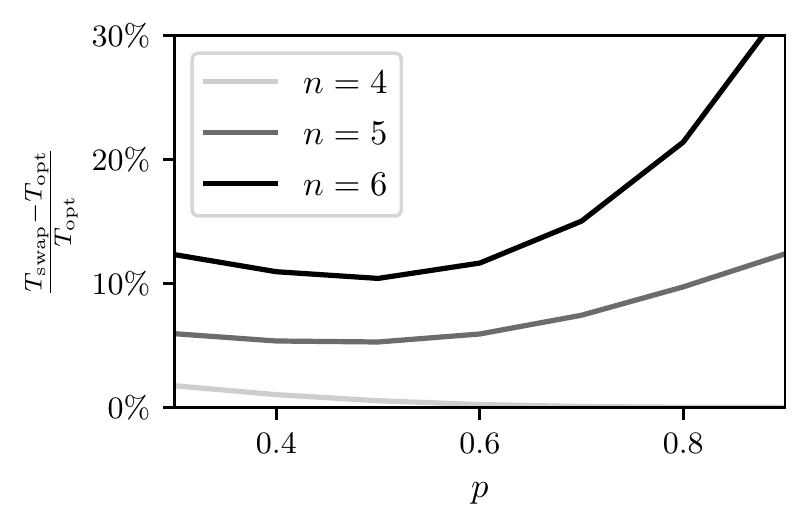}
         \caption{Probabilistic swaps ($p_\mathrm{s}=0.5$).}
	\label{fig.advantage-p-different-n-ps05}
     \end{subfigure}
     
	\caption{The advantage provided by an optimal policy over swap-asap is not always monotonic with $p$.
	Relative difference between the expected delivery times of an optimal policy, $T_\mathrm{opt}$, and the swap-asap policy, $T_\mathrm{swap}$, in an $n$-node chain with $t_\text{cut}=2$, for different values of $n$ and $p$.
	}
	\label{fig.advantage-p-different-n}
\end{figure}

In three-node chains, the swap-asap policy is always optimal since there is no reason to wait after both links have been generated.
As the number of nodes increases, policies have more degrees of freedom that can be adjusted to get an improvement over the swap-asap policy.
Figure~\ref{fig.advantage4} shows the advantage in expected delivery time of an optimal policy versus the swap-asap policy in four-node chains.
The swap-asap policy is no longer optimal, as it was in three-node chains, although the largest advantage observed is below $2\%$, meaning that the swap-asap policy is still close to optimal.
The advantage over swap-asap increases up to $30\%$ in five- and six-node chains, as shown in Figure~\ref{fig.advantage-n} (main text) and in Figure~\ref{fig.advantage-p-different-n}.

Figure~\ref{fig.advantage-p-different-n} shows the advantage of an optimal policy over swap-asap in terms of expected delivery time, versus $p$ and for different number of nodes.
When swaps are deterministic, the advantage is larger for smaller $p$.
The reason is that links are harder to generate as $p$ approaches zero, and therefore a fine-tuned policy that makes better use of those scarce resources is expected to be increasingly better than a greedy policy like swap-asap.
When swaps are probabilistic and $n>4$, the advantage is not monotonic in $p$ anymore, as can clearly be seen for $n=6$, $t_\text{cut}=2$, and $p_\text{s}=0.5$.
On the one hand, the advantage increases when $p$ approaches zero, due to swap-asap making an inefficient use of the links, which become a scarce resource.
On the other hand, when $p$ approaches one, the advantage also increases, due to the effect of full states, as discussed in the main text.

\vspace{20pt}

\section{Analysis of the actions of optimal policies}\label{app.swapasap-states}
\begin{figure}[t]
\captionsetup[subfigure]{justification=centering}
     \centering
     \begin{subfigure}[b]{0.37\textwidth}
         \centering
         \includegraphics[width=\textwidth]{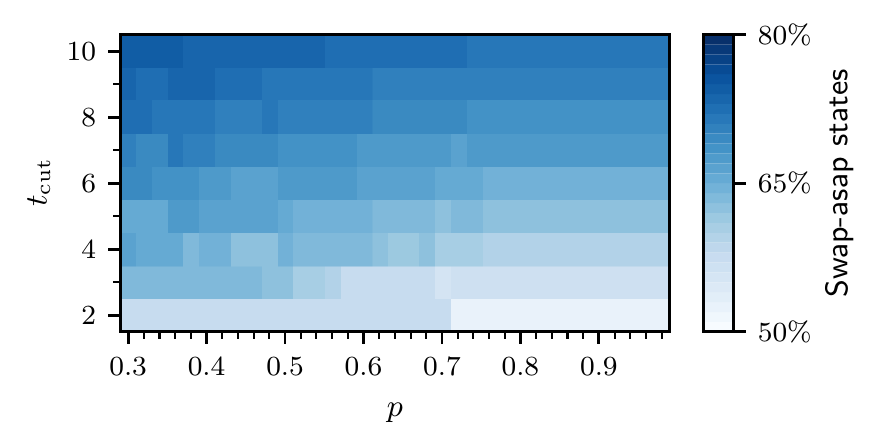}
         \caption{}
	\label{fig.states51swap}
     \end{subfigure}
     \hspace{40pt}
     \begin{subfigure}[b]{0.37\textwidth}
         \centering
         \includegraphics[width=\textwidth]{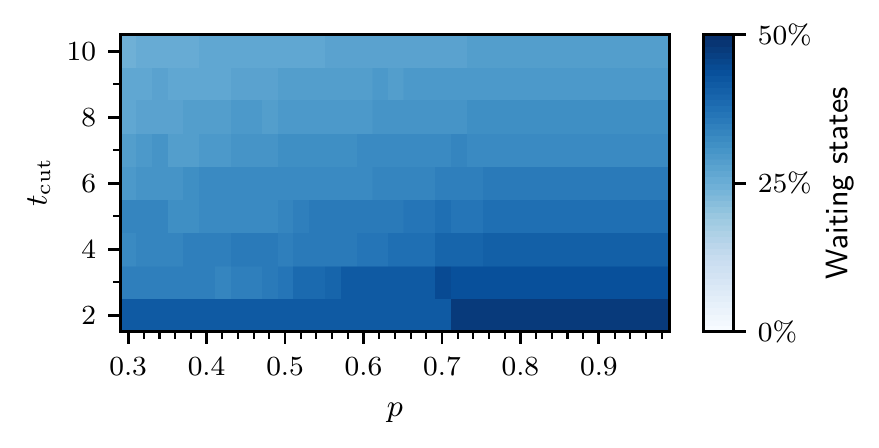}
         \caption{}
	\label{fig.states51wait}
     \end{subfigure}

     \begin{subfigure}[b]{0.37\textwidth}
         \centering
         \includegraphics[width=\textwidth]{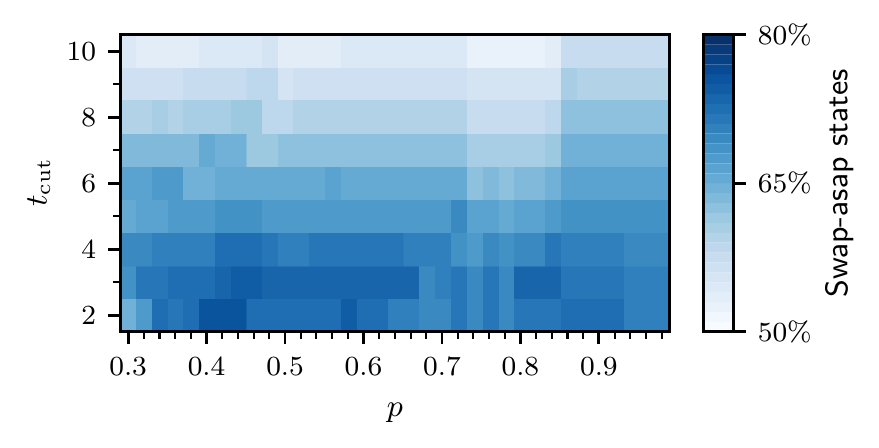}
         \caption{}
	\label{fig.states505swap}
     \end{subfigure}
     \hspace{40pt}
     \begin{subfigure}[b]{0.37\textwidth}
         \centering
         \includegraphics[width=\textwidth]{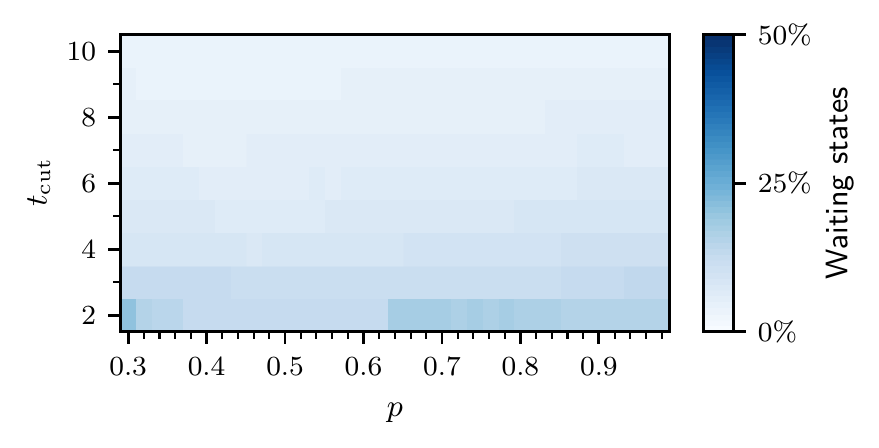}
         \caption{}
	\label{fig.states505wait}
     \end{subfigure}
     
	\caption{An optimal policy acts as the swap-asap policy in a large number of states.
	Percentage of states in which the optimal policy found by our solver decides to (a,c) perform all possible swaps or (b,d) not perform any swap, in a five-node repeater chain with (a-b) $p_\mathrm{s}=1$ or (c-d) $p_\mathrm{s}=0.5$. We only consider states in which at least one swap can be performed.}
	\label{fig.states5}
\end{figure}

Figure~\ref{fig.states5} shows the percentage of states in which an optimal policy decides to perform all possible swaps (acting as the swap-asap policy) or not perform any swap at all, in a five-node repeater chain. For this analysis, we only consider states in which at least one swap can be performed.
Although there seem to be some clear trends, these results cannot be used to determine how close the swap-asap policy is to being optimal, since:
\begin{enumerate}[label=(\roman*)]
	\item We found one of the possibly many optimal policies, so the swap-asap policy may be closer to a different optimal policy. This also explains why the plots in Figure~\ref{fig.states5} are not monotonic.
	\item Even if there is only one state in which two policies differ, this state could have a large impact on the expected delivery time, as explained in the example of full states in the main text.
\end{enumerate}

\begin{figure}[h]
\captionsetup[subfigure]{justification=centering}
     \centering
     \begin{subfigure}[b]{0.37\textwidth}
         \centering
         \includegraphics[width=\textwidth]{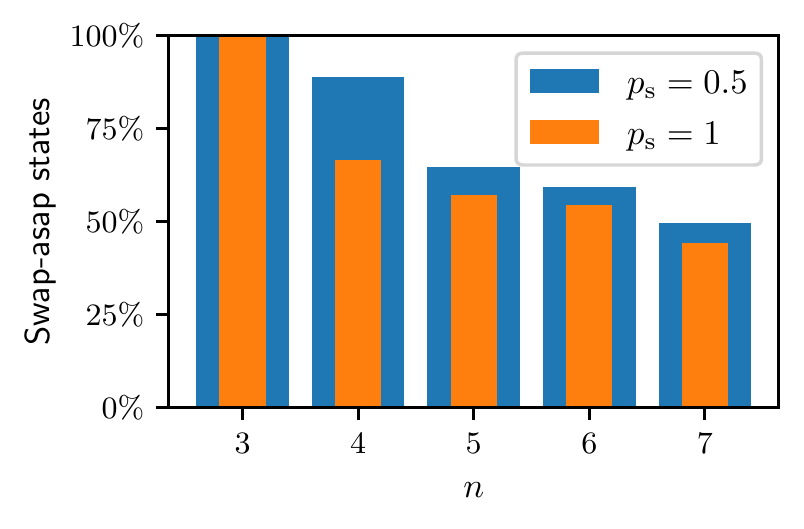}
	\vspace{-1.5\baselineskip}
         \caption{}
	\label{fig.swapasap-states-vs-n}
     \end{subfigure}
     \hspace{40pt}
     \begin{subfigure}[b]{0.37\textwidth}
         \centering
         \includegraphics[width=\textwidth]{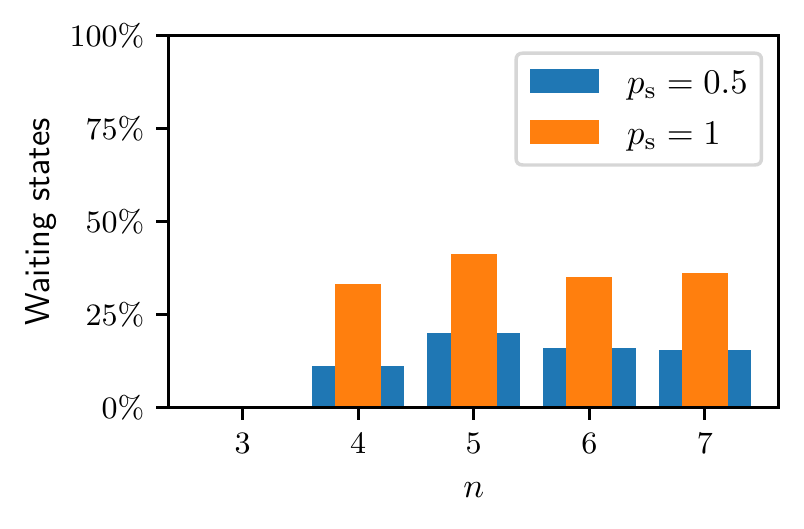}
	\vspace{-1.5\baselineskip}
         \caption{}
	\label{fig.waiting-states-vs-n}
     \end{subfigure}
     	\vspace{-0.5\baselineskip}
	\caption{The percentage of states in which the optimal policy acts as the swap-asap policy decreases in longer chains.
	Percentage of states in which the optimal policy found by our solver decides to (a) perform all possible swaps or (b) not perform any swap, in a repeater chain with $p=0.3$ and $\tc=2$. We only consider states in which at least one swap can be performed.}
	\label{fig.states-vs-n}
\end{figure}

In Figure~\ref{fig.states-vs-n} we plot the same quantities for increasing number of nodes.
The percentage of states in which the optimal policy decides to perform all possible swaps, acting as the swap-asap policy, decreases with increasing $n$.
This agrees with the fact that the advantage provided by an optimal policy in terms of expected delivery time over the swap-asap policy increases with increasing $n$, as shown in the main text.
However, the data from Figure~\ref{fig.states-vs-n} alone should not be used to draw any conclusions, since arguments ($i$) and ($ii$) also apply to these plots.

\vspace{20pt}

\section{Delivery time distribution}\label{app.delivery-time-distribution}
\begin{figure}[t]
\captionsetup[subfigure]{justification=centering}
     \centering
     \begin{subfigure}[b]{0.4\textwidth}
         \centering
         \includegraphics[width=\textwidth]{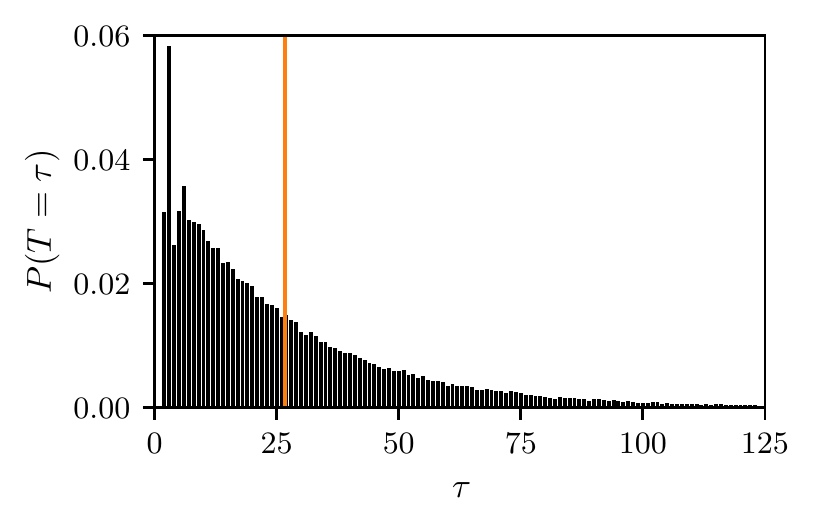}
         \caption{}
	\label{fig.deliv-time-distrib-p0.5}
     \end{subfigure}
     \hspace{40pt}
     \begin{subfigure}[b]{0.4\textwidth}
         \centering
         \includegraphics[width=\textwidth]{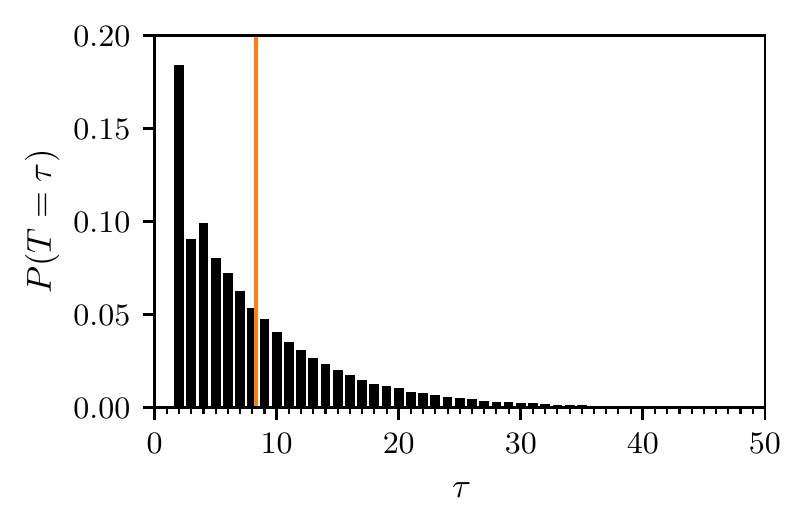}
         \caption{}
	\label{fig.deliv-time-distrib-p0.9}
     \end{subfigure}
	\caption{The delivery time distribution can be heavy-tailed.
	Delivery time distribution after simulating an optimal policy in a five-node repeater chain with $p_\text{s}=0.5$, $t_\text{cut}=2$, and (a) $p=0.5$ or (b) $p=0.9$. The number of samples is $10^5$.
	Solid orange lines correspond to the expected delivery time of the optimal policy.}
	\label{fig.deliv-time-distrib}
\end{figure}

Here, we show two examples of repeater chains in which the entanglement delivery time distribution is heavy-tailed.
Figure~\ref{fig.deliv-time-distrib} shows the delivery time distribution in a five-node chain with $p_\text{s}=0.5$, $t_\text{cut}=2$, and $p=0.5$ (Figure~\ref{fig.deliv-time-distrib-p0.5}) or $p=0.9$ (Figure~\ref{fig.deliv-time-distrib-p0.9}).
The results shown here have been calculated by repeatedly simulating the optimal policy in a repeater chain (source code available at \href{https://github.com/AlvaroGI/optimal-homogeneous-chain}{https://github.com/AlvaroGI/optimal-homogeneous-chain}).
As shown in the figure, the distribution is heavy-tailed for some combinations of parameters. In those cases, the average value does not provide an accurate description of the whole distribution.

\vspace{20pt}

\section{Expected time to reach an absorbing state}\label{app.hitting_time}
In this Appendix, we show that the expected time required to reach an absorbing state in a discrete Markov decision process (MDP) starting from state $\boldsymbol{s}$ and following policy $\pi$, $T_\pi(\boldsymbol{s})$, satisfies
\begin{equation*}
    T_\pi(\boldsymbol{s}) = 1 + \sum_{\boldsymbol{s}'\in\mathcal{S}} P(\boldsymbol{s}'|\boldsymbol{s},\pi) \cdot T_\pi(\boldsymbol{s}'),
\end{equation*}
where $\mathcal{S}$ is the state space and $P(\boldsymbol{s}'|\boldsymbol{s},\pi)$ is the probability of transition from state $\boldsymbol{s}$ to state $\boldsymbol{s}'$ when following policy $\pi$.
We also discuss the difference between deterministic and stochastic policies.

Let $t_\pi(\boldsymbol{s})$ be the time required to reach an absorbing state starting from state $\boldsymbol{s}$ in one realization of the process, and let
\begin{equation}\label{eq.app.T_s}
    T_\pi(\boldsymbol{s}) \equiv \mathbb{E}[t_\pi(\boldsymbol{s})] = \sum_{m=0}^{\infty} m \Pr[t_\pi(\boldsymbol{s})=m]
\end{equation}
be its expected value.
The time required to reach an absorbing state starting from $\boldsymbol{s}$ can be calculated as the time required to go from state $\boldsymbol{s}$ to any state $\boldsymbol{s}'$ plus the time required to go from $\boldsymbol{s}'$ to an absorbing state.
Since the Markov chain is discrete, we can write this as
\begin{equation}\label{eq.Pr_t_s}
    \Pr[t_\pi(\boldsymbol{s})=m] = \sum_{\boldsymbol{s}'\in\mathcal{S}} P(\boldsymbol{s}'|\boldsymbol{s},\pi) \Pr[t_\pi(\boldsymbol{s}')=m-1].
\end{equation}

The recursive relation for $T_\pi(\boldsymbol{s})$ can be derived as follows:
\begin{align}
    T_\pi(\boldsymbol{s}) \stackrel{a}{=}& \sum_{m=0}^{\infty} m \Pr[t_\pi(\boldsymbol{s})=m]\\
    \stackrel{b}{=}& \sum_{m=0}^{\infty} m \sum_{\boldsymbol{s}'\in\mathcal{S}} P(\boldsymbol{s}'|\boldsymbol{s},\pi) \Pr[t_\pi(\boldsymbol{s}')=m-1]\\
    =& \sum_{\boldsymbol{s}'\in\mathcal{S}} P(\boldsymbol{s}'|\boldsymbol{s},\pi) \sum_{m=0}^{\infty} m \Pr[t_\pi(\boldsymbol{s}')=m-1]\\
    =& \sum_{\boldsymbol{s}'\in\mathcal{S}} P(\boldsymbol{s}'|\boldsymbol{s},\pi) \sum_{m=-1}^{\infty} (m+1) \Pr[t_\pi(\boldsymbol{s}')=m]\\
    =& \sum_{\boldsymbol{s}'\in\mathcal{S}} P(\boldsymbol{s}'|\boldsymbol{s},\pi) \sum_{m=0}^{\infty} (m+1) \Pr[t_\pi(\boldsymbol{s}')=m]\\
    =& \sum_{\boldsymbol{s}'\in\mathcal{S}} P(\boldsymbol{s}'|\boldsymbol{s},\pi) \sum_{m=0}^{\infty} m \Pr[t_\pi(\boldsymbol{s}')=m]
    	+ \sum_{\boldsymbol{s}'\in\mathcal{S}} P(\boldsymbol{s}'|\boldsymbol{s},\pi) \sum_{m=0}^{\infty} \Pr[t_\pi(\boldsymbol{s}')=m]\\
    \stackrel{c}{=}& \sum_{\boldsymbol{s}'\in\mathcal{S}} P(\boldsymbol{s}'|\boldsymbol{s},\pi) \sum_{m=0}^{\infty} m \Pr[t_\pi(\boldsymbol{s}')=m]
       + 1\\
    \stackrel{d}{=}& \sum_{\boldsymbol{s}'\in\mathcal{S}} P(\boldsymbol{s}'|\boldsymbol{s},\pi) T_\pi(\boldsymbol{s}')
       + 1,
\end{align}
with the following steps:
\begin{enumerate}[label=\alph*.]
    \item We apply Equation (\ref{eq.app.T_s}).
    \item We apply Equation (\ref{eq.Pr_t_s}).
    \item We employ the normalization of the probability distributions: $\sum_{m=0}^{\infty} \Pr[t_\pi(\boldsymbol{s}')=m] = 1$ and $\sum_{\boldsymbol{s}'\in\mathcal{S}} P(\boldsymbol{s}'|\boldsymbol{s},\pi) = 1$.
    \item We use Equation (\ref{eq.app.T_s}) again.
\end{enumerate}

In the previous derivation, we have implicitly assumed that the policy is deterministic: at each state $\boldsymbol{s}$, the action chosen is always $\pi(\boldsymbol{s})$.
It can be shown that, in an MDP with a finite and countable set of states, there exists at least one optimal policy that is deterministic (see Section 2.3 from \cite{Szepesvari2010}). Therefore, since we are solving a finite MDP, we only need to consider deterministic policies. Optimal random policies can be built by combining several deterministic optimal policies, provided that there is more than one.

When considering stochastic policies, $\pi$ is no longer a mapping from a state to an action but a mapping from a state to a probability distribution over the action space.
The previous derivation remains valid for stochastic policies, although in that case the transition probabilities must be written as
\begin{equation}
	P(\boldsymbol{s}'|\boldsymbol{s},\pi) = \sum_{a\in\mathcal{A}} \pi(a|\boldsymbol{s}) P(\boldsymbol{s}'|\boldsymbol{s},a),
\end{equation}
where $\mathcal{A}$ is the action space and $\pi(a|\boldsymbol{s})$ is the probability of choosing action $a$ in state $\boldsymbol{s}$ when following policy $\pi$.

\vspace{20pt}

\section{Dynamic programming algorithms}\label{app.dp}
To find optimal policies, we formulate a Markov decision process that results in the Bellman equations, as explained in the main text.
These equations can be solved using a dynamic programming algorithm, such as value iteration and policy iteration.
Both algorithms start with arbitrary values of $T_\pi(\boldsymbol{s})$ (for some policy $\pi$ and $\forall \boldsymbol{s}\in\mathcal{S}$, where $\mathcal{S}$ is the state space) and they iteratively update the policy $\pi$ and the values $T_\pi(\boldsymbol{s})$, $\forall \boldsymbol{s}\in\mathcal{S}$.
The updated policy is guaranteed to converge to an optimal policy $\pi^*$ in a finite number of iterations in policy iteration and an infinite number of iterations in value iteration (see Sections 4.3 and 4.4 from \cite{Sutton2018}).
Note that there might be multiple optimal policies, although this approach finds only one of them.
In practice, the algorithms stop when the updated values differ by not more than some $\varepsilon>0$ from the values in the previous iteration. All our results have been calculated using $\varepsilon = 10^{-7}$.
In this work, we have applied both value iteration and policy iteration, which provided the same results (our specific implementations can be found at \href{https://github.com/AlvaroGI/optimal-homogeneous-chain}{https://github.com/AlvaroGI/optimal-homogeneous-chain}).
For a detailed explanation of both algorithms, see Sections 4.3 and 4.4 from~\cite{Sutton2018}.

In terms of computational cost, policy iteration is generally faster since less iterations are required. To the best of our knowledge, there are no known tight bounds on the number of iterations until convergence.
However, the computational complexity of a single iteration in policy iteration is $\mathcal{O}(|\mathcal{A}||\mathcal{S}|^2+|\mathcal{S}|^3)$, where $\mathcal{A}$ is the action space and $\mathcal{S}$ is the state space, which can be prohibitive for some combinations of parameters \cite{Kaelbling1996}.
The computational complexity of one iteration in value iteration is $\mathcal{O}(|\mathcal{A}||\mathcal{S}|^2)$ \cite{Kaelbling1996}.
In our problem, the complexity of each iteration increases exponentially with increasing number of nodes and polynomially with increasing cutoff time (see Appendix~\ref{app.scaling}), and the number of iterations increases with decreasing probability of entanglement generation and decreasing probability of successful swap, since the estimate of the values is worse when these probabilities are small.
Consequently, to study long chains with large cutoffs and small probabilities of successful entanglement generation and swap, one may need to employ approximate methods, such as deep reinforcement learning, which can find sub-optimal but good enough policies at a lower computational cost.

\vspace{20pt}

\section{Markov decision process example}\label{app.example}
Here we provide an example of how to formulate the Markov decision process (MDP) for a three-node repeater chain with cutoff $\tc=1$.
Specifically, we calculate each term in the Bellman equations, which can then be used to find an optimal policy, as explained in the main text.

\begin{figure}[h]
	\centering
	\includegraphics[width=0.6\textwidth]{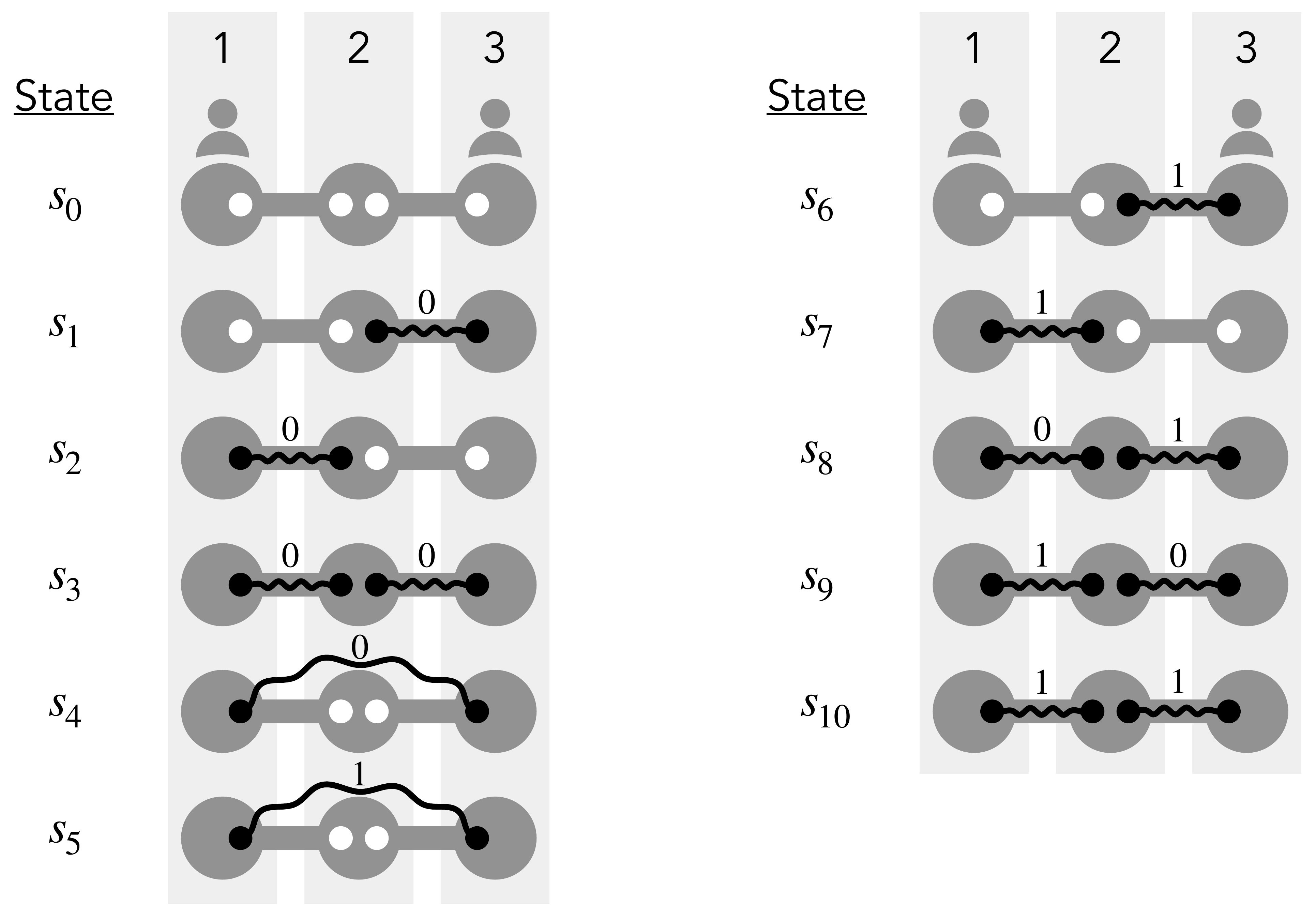}
	\caption{All possible states in a three-node repeater chain with cutoff $\tc=1$.
	Nodes are labeled 1 to 3 from left to right.}
	\label{fig.example3}
\end{figure}

We start by listing all the states in which the chain can be found. The sequence of events during each time slot is the following:
\begin{enumerate}
	\item First, the ages of all entangled links are increased by 1.
	\item Second, entanglement generation is attempted between every pair of neighbors with qubits available.
	\item Third, entanglement swaps can be performed.
	\item Lastly, links whose age is equal to $\tc$ are removed. End-to-end links are not removed.
\end{enumerate}
Since the cutoff is 1, the ages of all links are at most 1.
All possible states are listed in Figure~\ref{fig.example3}.

Let us now find the equation for $T_\pi(\boldsymbol{s}_0)$, the expected delivery time from state $\boldsymbol{s}_0$, which is given by
$$T_\pi(\boldsymbol{s}_0) = 1 + \sum_{\boldsymbol{s}'\in\mathcal{S}} P(\boldsymbol{s}'|\boldsymbol{s}_0,\pi) \cdot T_\pi(\boldsymbol{s}'),$$ as explained in the main text and derived in \ref{app.hitting_time}.
For clarity, let us abuse notation and write $T_i$ to denote $T_\pi(\boldsymbol{s}_i)$.
We can find each term by considering each possible scenario separately:
\begin{itemize}
	\item With probability $(1-p)^2$, no links are successfully generated and the state remains $\boldsymbol{s}_0$. Swaps and cutoffs do not apply to this state. This contributes with a term $(1-p)^2 T_0$.
	\item With probability $p(1-p)$, only one of the links is generated and the state becomes either $\boldsymbol{s}_1$ or $\boldsymbol{s}_2$. Swaps and cutoffs do not apply to these states. This contributes with $p(1-p)T_1 + p(1-p)T_2$.
	\item With probability $p^2$, both links are generated and the state becomes $\boldsymbol{s}_3$. Then, a swap can be performed and the last term splits into two contributions:
	\begin{itemize}
		\item If the policy decides to perform a swap in node 2, i.e., $\pi(\boldsymbol{s}_3)=\{2\}$, the state at the end of the time slot will be $\boldsymbol{s}_4$ if the swap is successful, and $\boldsymbol{s}_0$ if the swap fails. These scenarios contribute with $p^2 \mathds{1}_{\pi(\boldsymbol{s}_3)=\{2\}} \big(p_\mathrm{s} T_4 + (1-p_\mathrm{s}) T_0\big)$, where $\mathds{1}_{A}$ is the indicator function that takes value 1 if $A$ is true and value $0$ otherwise.
		\item If the policy decides to not perform the swap, i.e. $\pi(\boldsymbol{s}_3)=\emptyset$, the state remains $\boldsymbol{s}_3$. The contribution is then $p^2 \mathds{1}_{\pi(\boldsymbol{s}_3)=\emptyset} T_3$.
	\end{itemize}
\end{itemize}
Now, we can write all the terms above in a single equation:
\begin{equation}
	T_0 = 1 + (1-p)^2 T_0 + p(1-p)T_1 + p(1-p)T_2 + p^2 \mathds{1}_{\pi(\boldsymbol{s}_3)=\{2\}} \big(p_\mathrm{s} T_4 + (1-p_\mathrm{s}) T_0\big) + p^2 \mathds{1}_{\pi(\boldsymbol{s}_3)=\emptyset} T_3.
\end{equation}
Note that $T_4=T_5=0$, since $\boldsymbol{s}_4$ and $\boldsymbol{s}_5$ are absorbing states.
Rearranging terms, we obtain
\begin{equation}\label{eq.exT0}
	T_0 = 1 + \Big[(1-p)^2 + p^2 \mathds{1}_{\pi(\boldsymbol{s}_3)=\{2\}} (1-p_\mathrm{s})\Big] T_0
		+ \Big[ p(1-p)\Big] T_1 + \Big[ p(1-p) \Big] T_2
		+ \Big[ p^2 \mathds{1}_{\pi(\boldsymbol{s}_3)=\emptyset} \Big]T_3.
\end{equation}

Let us now find the equation for $\boldsymbol{s}_1$. At the beginning of the time slot, the age of the link is increased by 1 and the state becomes $\boldsymbol{s}_6$. After that:
\begin{itemize}
	\item With probability $(1-p)$, no links are successfully generated. Then, the only existing link is removed since it is 1 time slot old. This contributes with a term $(1-p) T_0$.
	\item With probability $p$, the remaining link is generated and the state becomes $\boldsymbol{s}_8$. Then, a swap can be performed:
	\begin{itemize}
		\item If the policy decides to perform a swap in node 2, i.e., $\pi(\boldsymbol{s}_8)=\{2\}$, the state at the end of the time slot will be $\boldsymbol{s}_5$ if the swap is successful, and $\boldsymbol{s}_0$ if the swap fails. These scenarios contribute with $p \mathds{1}_{\pi(\boldsymbol{s}_8)=\{2\}} \big(p_\mathrm{s} T_5 + (1-p_\mathrm{s}) T_0\big)$.
		\item If the policy decides to not perform the swap, i.e. $\pi(\boldsymbol{s}_8)=\emptyset$, the state remains $\boldsymbol{s}_8$, and the link with age 1 is removed afterwards. The state becomes $\boldsymbol{s}_2$ and the contribution is then $p \mathds{1}_{\pi(\boldsymbol{s}_8)=\emptyset} T_2$.
	\end{itemize}
\end{itemize}
Combining all terms, we obtain
\begin{equation}\label{eq.exT1}
	T_1 = 1 + \Big[(1-p) + p \mathds{1}_{\pi(\boldsymbol{s}_8)=\{2\}} (1-p_\mathrm{s}) \Big] T_0
		+ \Big[ p \mathds{1}_{\pi(\boldsymbol{s}_8)=\emptyset} \Big] T_2,
\end{equation}
where we have used that $T_5=0$.

Due to the symmetry of the problem,
\begin{equation}\label{eq.exT2}
	T_2=T_1,
\end{equation}
so we do not need to derive a new equation for $T_2$.

Lastly, we find the equation for $\boldsymbol{s}_3$. At the beginning of the time slot, the age of each link is increased by 1 and the state becomes $\boldsymbol{s}_{10}$, in which no more links can be generated. After that, a swap can be performed:
\begin{itemize}
	\item[--] If the policy decides to perform a swap in node 2, i.e., $\pi(\boldsymbol{s}_{10})=\{2\}$, the state at the end of the time slot will be $\boldsymbol{s}_5$ if the swap is successful, and $\boldsymbol{s}_0$ if the swap fails. These scenarios contribute with $\mathds{1}_{\pi(\boldsymbol{s}_{10})=\{2\}} \big(p_\mathrm{s} T_5 + (1-p_\mathrm{s}) T_0\big)$.
	\item[--] If the policy decides to not perform the swap, i.e. $\pi(\boldsymbol{s}_{10})=\emptyset$, the state remains $\boldsymbol{s}_{10}$, and both links are removed afterwards, when cutoffs are applied. The state becomes $\boldsymbol{s}_0$ and the contribution is then $\mathds{1}_{\pi(\boldsymbol{s}_{10})=\emptyset} T_0$.
\end{itemize}
The equation then reads
\begin{equation}\label{eq.exT3}
	T_3 = 1 + \Big[ \mathds{1}_{\pi(\boldsymbol{s}_{10})=\{2\}} (1-p_\mathrm{s}) + \mathds{1}_{\pi(\boldsymbol{s}_{10})=\emptyset} \Big] T_0,
\end{equation}
where we have used that $T_5=0$.

We can write Equations (\ref{eq.exT0}), (\ref{eq.exT1}), (\ref{eq.exT2}), and (\ref{eq.exT3}) as
\begin{equation}\label{eq.exTsystem}
\begin{cases}
	\vspace{3pt}
	T_0 = 1 + \Big[(1-p)^2 + p^2 \mathds{1}_{\pi(\boldsymbol{s}_3)=\{2\}} (1-p_\mathrm{s})\Big] T_0
		+ 2\Big[ p(1-p)\Big] T_1
		+ \Big[ p^2 \mathds{1}_{\pi(\boldsymbol{s}_3)=\emptyset} \Big]T_3,\\
	\vspace{3pt}
	T_1 = 1 + \Big[(1-p) + p \mathds{1}_{\pi(\boldsymbol{s}_8)=\{2\}} (1-p_\mathrm{s}) \Big] T_0
		+ \Big[ p \mathds{1}_{\pi(\boldsymbol{s}_8)=\emptyset} \Big] T_1,\\
	T_3 = 1 + \Big[ \mathds{1}_{\pi(\boldsymbol{s}_{10})=\{2\}} (1-p_\mathrm{s}) + \mathds{1}_{\pi(\boldsymbol{s}_{10})=\emptyset} \Big] T_0.
\end{cases}
\end{equation}
An optimal policy $\pi^*$ can be found by minimizing $T_0$, $T_1$, and $T_3$ in this system of equations. This can be done, e.g., using iterative algorithms such as value and policy iteration, as discussed in the main text.
In this case, it can be shown that the swap-asap policy is optimal, i.e., $\pi^*(\boldsymbol{s}_3)=\pi^*(\boldsymbol{s}_8)=\pi^*(\boldsymbol{s}_{10})=\{2\}$. This also makes sense intuitively: once both links are generated, waiting provides no advantage in terms of delivery time over performing the swap immediately.
For this policy, the system of equations becomes
\begin{equation}\label{eq.exTsystem}
\begin{cases}
	\vspace{3pt}
	T_0 = 1 + \Big[(1-p)^2 + p^2 (1-p_\mathrm{s})\Big] T_0
		+ 2\Big[ p(1-p)\Big] T_1,\\
	\vspace{3pt}
	T_1 = 1 + \Big[(1 - p p_\mathrm{s} ) \Big] T_0,\\
	T_3 = 1 + \Big[(1-p_\mathrm{s}) \Big] T_0,
\end{cases}
\end{equation}
which yields an optimal expected delivery time of
$$T_0 = \frac{1+2p(1-p)}{1 - (1-p)^2 - p^2(1-p_\mathrm{s}) - 2p(1-p)(1-pp_\mathrm{s})}.$$

As a final remark, note that the expected delivery times from states $\boldsymbol{s}_6$ to $\boldsymbol{s}_{10}$ were not necessary to compute $T_0$. In fact, states $\boldsymbol{s}_6$ to $\boldsymbol{s}_{10}$ cannot exist at the beginning of a time slot, since the links that exist at the beginning of a time slot are always younger than $\tc$ (i.e., their age is 0) or are end-to-end links. This is the reason why we do not need to optimize over $T_6$ to $T_{10}$.

\vspace{20pt}

\section{Scaling of the number of states}\label{app.scaling}
In this Appendix, we find a lower bound to the number of states in the Markov decision process discussed in the main text, and show that it scales as $\Omega\big((\tc)^{n-2}\big)$. Then, we compare this lower bound to the exact number of states for some combinations of parameters.

We start by calculating the lower bound.
Let us define $\mathcal{S}(l)$ as the set of states in which only $l$ entangled links are present. By definition,
\begin{equation*}
    \mathcal{S} = \bigcup_{l} \mathcal{S}(l).
\end{equation*}
The sets $\mathcal{S}(l)$ do not overlap. Therefore,
\begin{equation}\label{eq.S_sum_Sl}
    |\mathcal{S}| = \sum_{l} |\mathcal{S}(l)|.
\end{equation}

The first term is given by
\begin{equation}\label{eq.S0}
    |\mathcal{S}(0)|=1,
\end{equation}
since there is only one state without any entangled links.

Since any two nodes could potentially share an entangled link, there are
$$k = {n\choose2}-1 = \frac{n^2-n-2}{2}$$
possible links in a repeater chain with $n$ nodes (note that we subtract 1 from the combinatorial number  because there is no need to represent end-to-end links).
When one of those links exists, its age can be anything from 0 to $\tc$. Therefore, the total number of different states with only one link is given by
\begin{equation}\label{eq.S1}
    |\mathcal{S}(1)| = \frac{n^2-n-2}{2} (\tc+1) \geq \frac{n^2-n-2}{2} \tc.
\end{equation}

Let us now consider a chain with two links where both are connected to the same node $i$. From node $i$ towards one end of the chain, there are $i-1$ nodes. From $i$ towards the other end of the chain, there are $n-i$ nodes.
Then, the number of states with two links where both are connected to node $i$ is given by
\begin{equation*}
    |\mathcal{S}_i(2)| = (i-1)(n-i)(\tc+1)^2,
\end{equation*}
where the last factor accounts for all possible ages of both links.
We can find a lower bound to $|\mathcal{S}(2)|$ by considering only the states in which both links are connected to the same node $i$, i.e.,
\begin{equation}\label{eq.S2}
\begin{split}
    |\mathcal{S}(2)| &\geq \sum_{i=2}^{n-1} |\mathcal{S}_i(2)|\\
    &= \sum_{i=2}^{n-1} (i-1)(n-i)(\tc+1)^2\\
    &= \sum_{j=1}^{n-2} j(n-j-1)(\tc+1)^2\\
    &= (\tc+1)^2 \sum_{j=1}^{n-2} \big( -j^2+(n-1)j \big)\\
    &= (\tc+1)^2 \frac{n(n-1)(n-2)}{6}\\
    &\geq \frac{n(n-1)(n-2)}{6} \tc^2,
\end{split}
\end{equation}
where we used the identities
$\sum_{k=1}^{m}k = \frac{m(m+1)}{2}$
and
$\sum_{k=1}^{m}k^2 = \frac{m(m+1)(2m+1)}{6}$ to simplify the sum.
\\

Next, we compute a lower bound for $|\mathcal{S}(l)|$, $l>2$. Let us consider only states with $l$ adjacent links, i.e., states that have $l-1$ nodes that hold 2 entangled links. The number of such states is lower bounded by the different ways in which we can pick those $l-1$ nodes from a total of $n-2$ nodes (end-nodes cannot have two links) and the possible ages of the $l$ links.
Therefore, we can bound $|\mathcal{S}(l)|$ as follows:
\begin{equation}\label{eq.Sl}
    |\mathcal{S}(l)| \geq {n-2 \choose l-1} (\tc+1)^l \geq {n-2 \choose l-1} \tc^l, \;\; l>2.
\end{equation}

Finally, using Equations (\ref{eq.S_sum_Sl}) to (\ref{eq.Sl}), we find a lower bound for $|\mathcal{S}|$:
\begin{align*}
    |\mathcal{S}| =&\; \sum_{l=0}^{2} |\mathcal{S}(l)| + \sum_{l=3}^{n-1} |\mathcal{S}(l)|\\
    \geq&\; \sum_{l=0}^{2} |\mathcal{S}(l)|
         + \sum_{l=3}^{n-1} {n-2 \choose l-1} \tc^{l}\\
    \geq&\; \sum_{l=0}^{2} |\mathcal{S}(l)|
        + \tc \sum_{k=0}^{n-2} {n-2 \choose k} \tc^{k}
        - \tc - (n-2)\tc^2\\
    \stackrel{a}{\geq}&\; \sum_{l=0}^{2} |\mathcal{S}(l)|
        + \tc(\tc+1)^{n-2} - \tc - (n-2)\tc^2\\
    \geq&\; \sum_{l=0}^{2} |\mathcal{S}(l)|
        + \tc^{n-1} - \tc - (n-2)\tc^2\\
    \geq&\; 1 + \frac{n^2-n-2}{2} \tc
	+ \frac{n(n-1)(n-2)}{6} \tc^2
	+ \tc^{n-1} - \tc - (n-2)\tc^2\\
\end{align*}
where, in step $a$, we have used the binomial sum: $$\sum_{k=0}^n {n\choose k}x^k = (1+x)^n.$$
After some algebra, we find
\begin{equation}\label{eq.Slowerbound}
	|\mathcal{S}| \geq 1 + \frac{n^2-n-4}{2} \tc + \frac{(n^2-n-6)(n-2)}{6} \tc^2 + \tc^{n-1}.
\end{equation}

From the previous result, we conclude that the scaling of the number of states is
$$|\mathcal{S}| = \Omega\big((\tc)^{n-1}\big).$$

\begin{figure}[t]
\captionsetup[subfigure]{justification=centering}
     \centering
     \begin{subfigure}[b]{0.4\textwidth}
         \centering
         \includegraphics[width=\textwidth]{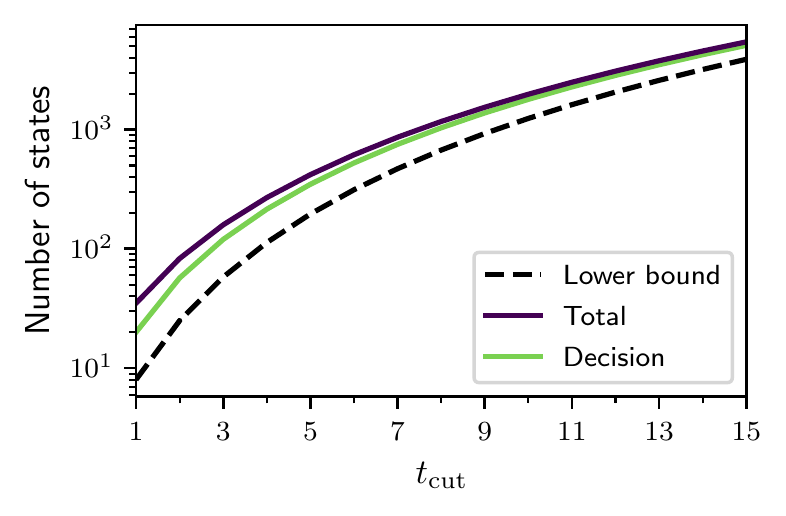}
         \caption{}
         \label{fig.scaling_tc}
     \end{subfigure}
     \begin{subfigure}[b]{0.4\textwidth}
         \centering
         \includegraphics[width=\textwidth]{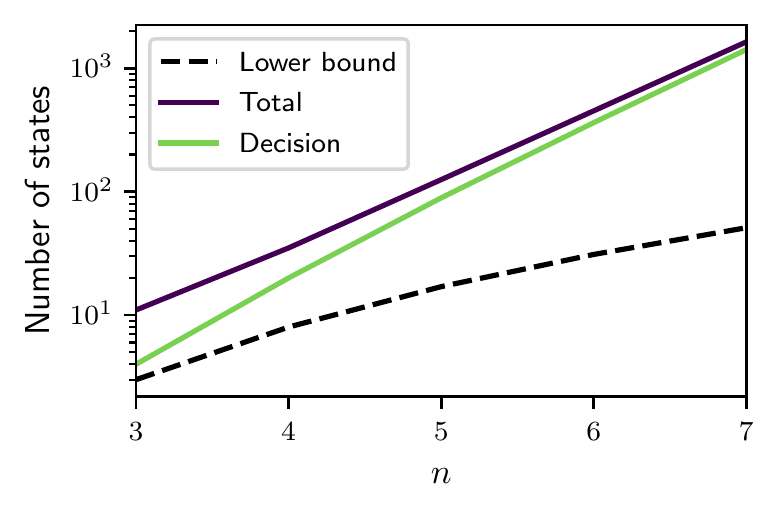}
         \caption{}
         \label{fig.scaling_n}
     \end{subfigure}
     
	\caption{The number of states scales at least exponentially with increasing $n$ and polynomially with increasing $t_\mathrm{cut}$.
	(a) Number of states versus the cutoff time in a four-node chain, and (b) versus the number of nodes in a chain with cutoff $t_\mathrm{cut}=1$.
	Solid lines correspond to the number of states found by our policy iteration algorithm (note that the number of states only depends on $n$ and $\tc$).
	The purple solid line is the total number of states and the green line is the number of states in which a decision can be made (i.e., states in which at least one swap can be performed).
	The dashed line corresponds to the lower bound (\ref{eq.Slowerbound}) to the total number of states.}
	\label{fig.scaling}
\end{figure}

Figure~\ref{fig.scaling} shows the exact number of states versus the cutoff time and the number of nodes, together with the lower bound (\ref{eq.Slowerbound}).
In these plots, the exact number of states corresponds to the size of the state space explored by our policy iteration algorithm.

\vspace{20pt}

\section{Calculation of transition probabilities with state bunching}\label{app.environment}
In this Appendix, we explain how to simplify the calculation of transition probabilities with a technique that we call \textit{state bunching}.
This method takes advantage of state symmetries to reduce the number of equations and variables in the Bellman equations. Let us discuss how it works.

\begin{definition}
The mirrored version of a state is obtained by relabeling the nodes in reverse order. We denote the mirrored version of $\boldsymbol{s}$ as $\text{mirror}(\boldsymbol{s})$.
\end{definition}

\begin{figure}[ht!]
\captionsetup[subfigure]{justification=centering}
    \centering
    \subfloat[An example state.\vspace{10pt}]{\includegraphics[width=0.4\textwidth]{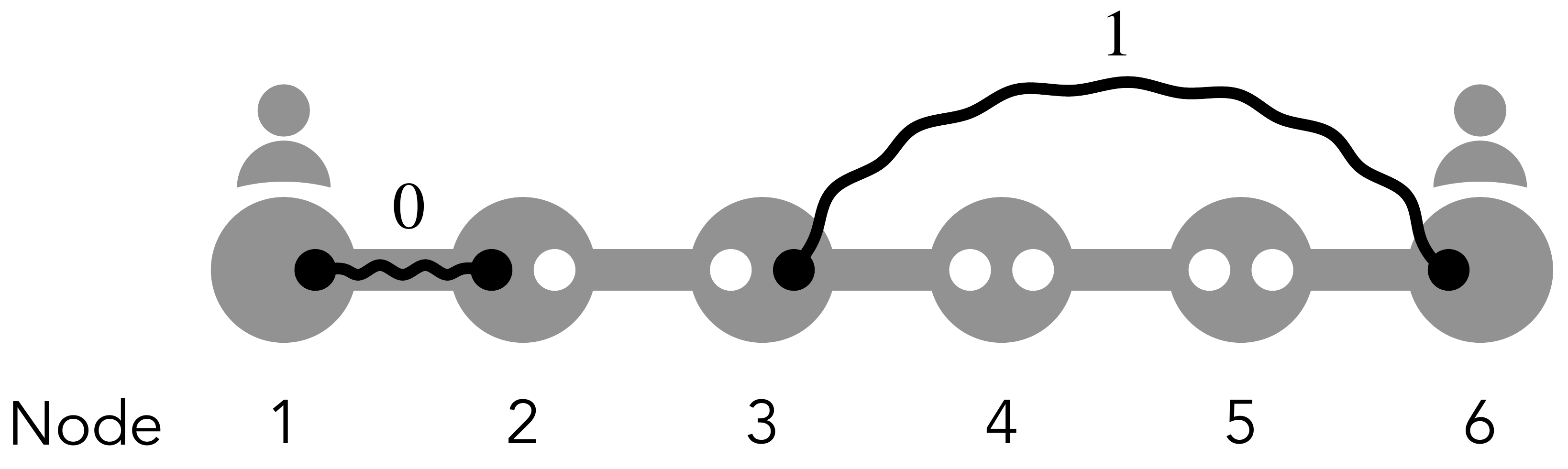}}\\
    \subfloat[The mirrored version of the state above.]{\includegraphics[width=0.4\textwidth]{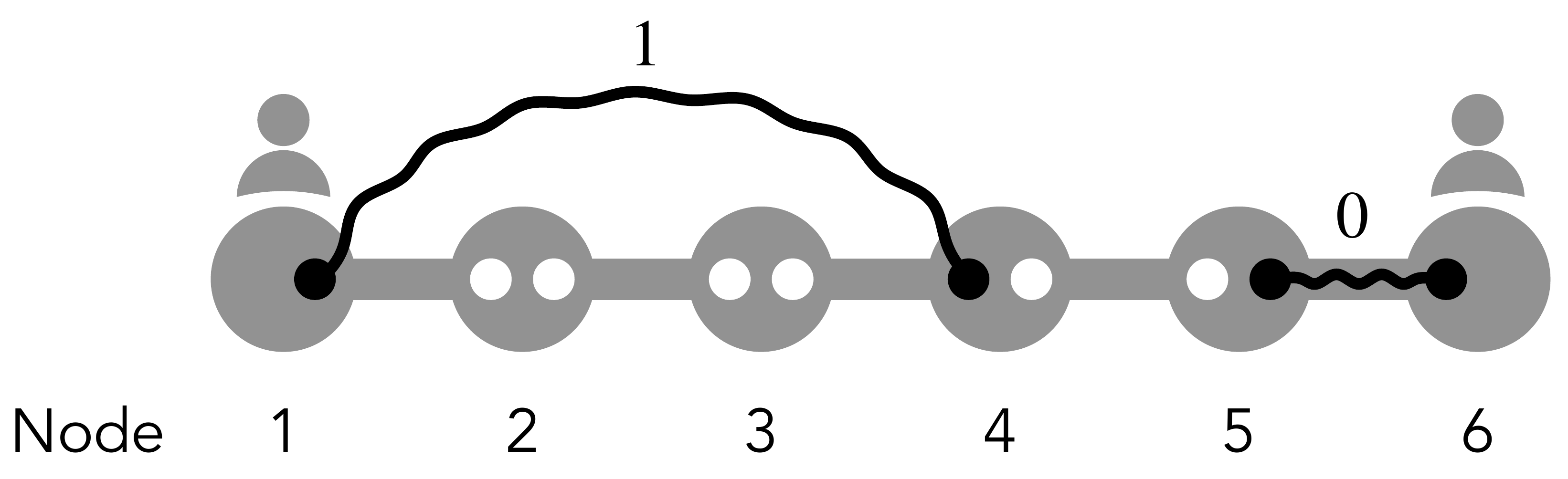}}
    \caption{Example of a non-symmetric state and its mirrored version in a six-node repeater chain. Solid black lines represent entangled links, with occupied qubits in black and free qubits in white. The number above each entangled link corresponds to its age.}
    \label{fig:state_and_mirror}
\end{figure}

\begin{definition}
A symmetric (sym) state $\boldsymbol{s}$ is one that is identical to its mirrored version, i.e., $\text{mirror}(\boldsymbol{s})=\boldsymbol{s}$. All other states are non-symmetric (non-sym).
\end{definition}
\begin{lemma}\label{lemma.nonsym}
	Every non-sym state $\boldsymbol{s}$ has a mirrored version $\boldsymbol{\tilde s}$ that is different from $\boldsymbol{s}$, i.e.,
	$$\forall \boldsymbol{s}\in \mathcal{S} \;\,\text{s.t.}\;\, \boldsymbol{s}\neq\text{mirror}(\boldsymbol{s}),\;\, \exists \; \boldsymbol{\tilde s} \in \mathcal{S} \;\,\text{s.t.}\;\, \text{mirror}(\boldsymbol{s}) = \boldsymbol{\tilde s} \;\,\text{and}\;\, \boldsymbol{s}\neq \boldsymbol{\tilde s}.$$
\end{lemma}
The lemma above stems from the definition of non-sym state.
Figure~\ref{fig:state_and_mirror} presents an example of non-sym state and its mirrored version.

Let $\mathcal{S}$ be the set of states, $\mathcal{S}_\text{s}$ be the set of sym states, and $\mathcal{S}_\text{ns}$ be the set of non-sym states.
Due to symmetry, the expected delivery times of a non-sym state and its mirrored version are equal, i.e., $T_\pi(\boldsymbol{s}) = T_\pi(\text{mirror}(\boldsymbol{s}))$, for any policy $\pi$ and any $\boldsymbol{s}\in \mathcal{S}_\text{ns}$.
Hence, we can remove half of the non-sym states from the Bellman equations to reduce the computational complexity, by following these steps:
\begin{enumerate}
	\item Group all non-sym states in two sets, $\mathcal{S}_\text{1}$ and $\mathcal{S}_\text{2}$, such that neither set contains a state and its mirrored version, i.e., define $\mathcal{S}_\text{1}$ and $\mathcal{S}_\text{2}$ such that
	$$\mathcal{S}_1 \cup \mathcal{S}_2 = \mathcal{S}_\text{ns}
	\;\;\text{and}\;\; \boldsymbol{s}_1\neq\text{mirror}(\boldsymbol{s}_1'), \forall \boldsymbol{s}_1,\boldsymbol{s}_1'\in\mathcal{S}_1
	\;\;\text{and}\;\; \boldsymbol{s}_2\neq\text{mirror}(\boldsymbol{s}_2'), \forall \boldsymbol{s}_2,\boldsymbol{s}_2'\in\mathcal{S}_2.$$
	
	\item Replace $T_\pi(\boldsymbol{s}_2)$ by $T_\pi(\text{mirror}(\boldsymbol{s}_2))$, for every $\boldsymbol{s}_2\in\mathcal{S}_2$ in the Bellman equations. This can be done since $T_\pi(\boldsymbol{s}_2) = T_\pi(\text{mirror}(\boldsymbol{s}_2))$.
	Note that $\text{mirror}(\boldsymbol{s}_2)\in\mathcal{S}_1$.
	
	\item The Bellman equations are now given by
	$$T_\pi(\boldsymbol{s}) = 1 
		+ \sum_{\boldsymbol{s}'\in\mathcal{S}_\text{s}\cup\mathcal{S}_1} P(\boldsymbol{s}'|\boldsymbol{s},\pi) \cdot T_\pi(\boldsymbol{s}')
		+ \sum_{\boldsymbol{s}'\in\mathcal{S}_2} P(\boldsymbol{s}'|\boldsymbol{s},\pi) \cdot T_\pi(\text{mirror}(\boldsymbol{s}')), \; \forall \boldsymbol{s}\in\mathcal{S}_\text{s}\cup\mathcal{S}_1,$$
	which can also be written as
	\begin{equation}\label{eq.Bellmanbunching}
		T_\pi(\boldsymbol{s}) = 1 
		+ \sum_{\boldsymbol{s}'\in\mathcal{S}_\text{s}\cup\mathcal{S}_1} \tilde P(\boldsymbol{s}'|\boldsymbol{s},\pi) \cdot T_\pi(\boldsymbol{s}'),
		\; \forall \boldsymbol{s}\in\mathcal{S}_\text{s}\cup\mathcal{S}_1,
	\end{equation}
	where
	\begin{equation*}
	\tilde P(\boldsymbol{s}'|\boldsymbol{s},\pi) = 
	\begin{cases}
		P(\boldsymbol{s}'|\boldsymbol{s},\pi), \;\;\text{if}\;\; \boldsymbol{s}'\in\mathcal{S}_\text{s} \\
		P(\boldsymbol{s}'|\boldsymbol{s},\pi) + P(\text{mirror}(\boldsymbol{s}')|\boldsymbol{s},\pi), \;\;\text{if}\;\; \boldsymbol{s}'\in\mathcal{S}_1
	\end{cases}.
	\end{equation*}
\end{enumerate}

When using state bunching, we need to optimize a system of $|\mathcal{S}_\text{s}|+|\mathcal{S}_1| < |\mathcal{S}|$ variables, as can be seen from (\ref{eq.Bellmanbunching}).

\begin{figure}[th!]
    \centering
    \includegraphics[width=0.3\textwidth]{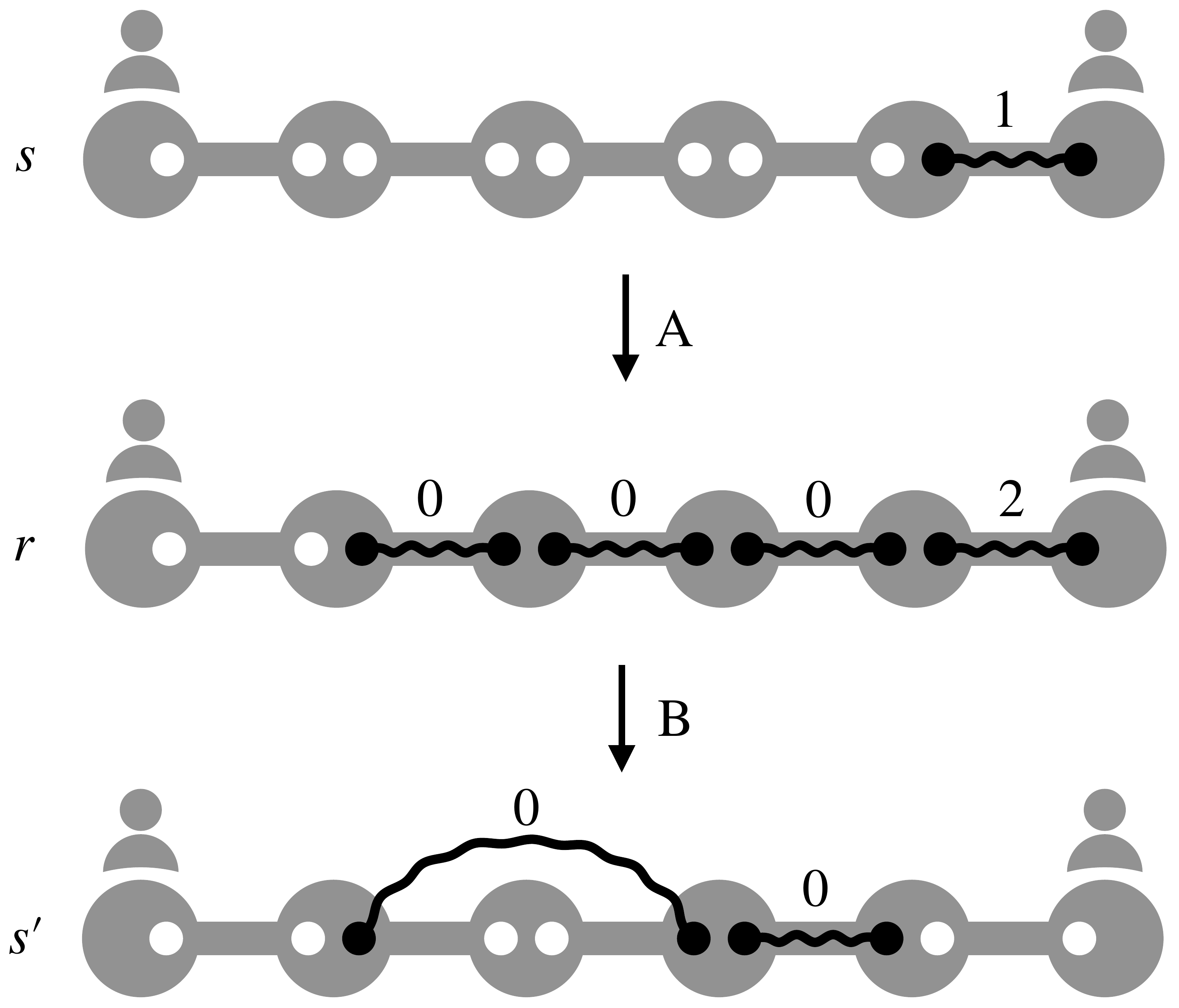}
    \caption{Example of state evolution within a single time slot for a chain with six nodes and cutoff $t_\text{cut}=2$.
    The initial state is $\boldsymbol{s}$. During part A of the time slot, we update the age of every link and attempt entanglement generation, leading to an intermediate state $\boldsymbol{r}$.
    During part B of the time slot, swaps are performed and cutoffs are applied, leading to a final state $\boldsymbol{s}'$.}
    \label{fig:state_evolutionn}
\end{figure}

Let us now assume that swaps are deterministic.
With this assumption, we provide an analytical framework to simplify the calculation of the transition probabilities.
Since there are several events happening in a single time slot, it is convenient to split $P$ into two contributions.
Each time slot can be divided in two parts:
\begin{enumerate}[label=\Alph*.]
	\item The age of every link is updated (i.e., we add one to every link age), and then entanglement generation is attempted wherever qubits are available.
	\item Swaps are performed according to action $a$, and then cutoffs are applied.
\end{enumerate}
\begin{definition}
	An intermediate state $\boldsymbol{r}$ is a state that exists between parts A and B of a time slot.
\end{definition}
Figure~\ref{fig:state_evolutionn} presents an example of this division of the time slot for a six-node chain with cutoff $t_\text{cut}=2$.
Let us define $P_\text{A}(\boldsymbol{r}|\boldsymbol{s})$ as the probability that an intermediate state $\boldsymbol{r}$ is produced from an initial state $\boldsymbol{s}$ after the first part of the time slot. Note that $P_\text{A}$ does not depend on any action, since swaps are performed after part A.
Similarly, let us define $P_\text{B}(\boldsymbol{s}'|\boldsymbol{r},a)$ as the probability that the state at the end of the time slot is $\boldsymbol{s}'$ given that $\boldsymbol{r}$ was the intermediate state and action $a$ was performed.
We can write the transition probability from $\boldsymbol{s}$ to $\boldsymbol{s}'$ as follows:
$$P(\boldsymbol{s}'|\boldsymbol{s},a) = \sum_{\boldsymbol{r}\in\mathcal{S}} P_\text{B}(\boldsymbol{s}'|\boldsymbol{r},a) P_\text{A}(\boldsymbol{r}|\boldsymbol{s}).$$
Note that we only need to consider $\boldsymbol{s}\in \mathcal{S}_\text{s}\cup\mathcal{S}_1$ to solve (\ref{eq.Bellmanbunching}), while $\boldsymbol{s}',\boldsymbol{r} \in \mathcal{S}_\text{s}\cup\mathcal{S}_1\cup\mathcal{S}_2$.
Next, we show that the calculation of $P$ can be further simplified by ignoring some terms in the sum while adding some multipliers in other terms.
We need the following definitions:
\begin{definition}
	The \textit{label of an entangled link} that is shared between nodes $i$ and $j$ is a tuple $(i,j)$. Note that the label $(i,j)$ is equivalent to $(j,i)$.
\end{definition}

\begin{definition}
	Symmetry-preserving (SP) links of a repeater chain are those entangled links that retain their original labels in the mirrored version of a state. We denote all other links as non-symmetry-preserving (NSP).
\end{definition}

\begin{definition}
	The mirrored link of an NSP link with label $(i,j)$ and age $g$ is a link with label $(n-j+1,n-i+1)$ and age $g$.
\end{definition}

If the transition during part A of the time slot is from a sym state $\boldsymbol{s}$ to a non-sym state $\boldsymbol{r}\in\mathcal{S}_1$, then a multiplier of two is necessary to omit states in $\mathcal{S}_2$. The reason is as follows: to transition to a non-sym state via entanglement generation, the chain must have generated an NSP link, and at the same time it must not have generated its mirrored link (the mirrored link can always be generated, since we start from a sym state). Note that generating an SP link does not affect the initial state's symmetry. Thus, the intermediate state $\boldsymbol{r}$ contains a new NSP link; however, had the chain generated its mirrored link instead, it would have generated $\text{mirror}(\boldsymbol{r})\in\mathcal{S}_2$.
Note that the probability of generating the NSP link is equal to that of generating its mirrored link, i.e., the transition probabilities to $\boldsymbol{r}$ and $\text{mirror}(\boldsymbol{r})$ are the same.
Hence, the transition from $\boldsymbol{s}$ to $\text{mirror}(\boldsymbol{r})\in\mathcal{S}_2$ can be captured by introducing a factor of two in $P_\text{A}(\boldsymbol{r}|\boldsymbol{s})$.
This way, we do not need to consider intermediate states $\boldsymbol{r}$ in $\mathcal{S}_2$ when $\boldsymbol{s}$ is a sym state.

Suppose now that the transition is from a sym state $\boldsymbol{s}$ to another sym state $\boldsymbol{r}$.
Since both states are sym, the transition probability $P_\text{A}(\boldsymbol{r}|\boldsymbol{s})$ is not affected when removing some non-sym states in state bunching.

Next, we note that transitions from a non-sym state $\boldsymbol{s}$ to a sym state $\boldsymbol{r}$ are not possible: if the initial state is non-sym, then it contains at least one NSP link without also containing its mirrored link. Such links would thus also be present in the intermediate state $\boldsymbol{r}$, with incremented ages, so that even if their mirrored links are generated, the new state would still be non-sym due to the age mismatch. It follows, then, that a non-sym state $\boldsymbol{s}\in\mathcal{S}_1$ may only transition to another non-sym state $\boldsymbol{r}\in\mathcal{S}_1$, and due to the presence of at least one NSP link with absence of its mirrored link, such a transition may only generate $\boldsymbol{r}\in\mathcal{S}_1$ and not $\text{mirror}(\boldsymbol{r})\in\mathcal{S}_2$ (here, we assumed that $\boldsymbol{r}$ was placed in $\mathcal{S}_1$ when assigning non-sym states to each set; if this was not the case, we only need to exchange $\boldsymbol{r}$ and $\text{mirror}(\boldsymbol{r})$ to the opposite set).
Thus, if both $\boldsymbol{s}$ and $\boldsymbol{r}$ are non-sym states, $P_\text{A}(\boldsymbol{r}|\boldsymbol{s})$ is unaffected by state bunching. Figure~\ref{fig:state_evolutionn} depicts an example of such a transition.

Next, we shift our focus to $P_\text{B}(\boldsymbol{s}'|\boldsymbol{r},a)$, i.e., transitions that occur as a result of (deterministic) entanglement swaps, followed by link expirations.
During part B of the time slot, state $\boldsymbol{r}\in\mathcal{S}_\text{s}\cup\mathcal{S}_1$ may be taken to some state $\boldsymbol{s}'$ or $\text{mirror}(\boldsymbol{s}')$, depending on the combination of swaps.
Take, for instance, the transition in Figure~\ref{fig:state_evolutionn}: if the swap had been performed at the fourth node instead of the third, then the chain would have generated the mirrored version of the final state.
Nevertheless, the chain must ultimately choose only one swapping combination at this stage. Since swaps and cutoffs are deterministic, either $\boldsymbol{s}'$ or $\text{mirror}(\boldsymbol{s}')$ (or none of them) is obtained deterministically.
Therefore, there is no need to update $P_\text{B}(\boldsymbol{s}'|\boldsymbol{r},a)$ when using state bunching.

Finally, the transition probabilities can be written as
\begin{equation}
\begin{split}
	P(\boldsymbol{s}'|\boldsymbol{s},a) &= \sum_{\boldsymbol{r}\in\mathcal{S}_\text{s}} P_\text{B}(\boldsymbol{s}'|\boldsymbol{r},a) P_\text{A}(\boldsymbol{r}|\boldsymbol{s})
			+ \sum_{\boldsymbol{r}\in\mathcal{S}_1} P_\text{B}(\boldsymbol{s}'|\boldsymbol{r},a)\cdot 2\cdot P_\text{A}(\boldsymbol{r}|\boldsymbol{s}),
			\;\; \text{if} \;\; \boldsymbol{s}\in\mathcal{S}_\text{s},\\
	P(\boldsymbol{s}'|\boldsymbol{s},a) &= \sum_{\boldsymbol{r}\in\mathcal{S}_1} P_\text{B}(\boldsymbol{s}'|\boldsymbol{r},a) P_\text{A}(\boldsymbol{r}|\boldsymbol{s}),
			\;\; \text{if} \;\; \boldsymbol{s}\in\mathcal{S}_\text{1},
\end{split}
\end{equation}
where we have removed all intermediate states from $\mathcal{S}_2$.

\end{document}